\documentclass[prb,twocolumn,showpacs,groupedaddress,amsmath,amssymb]{revtex4-1}

\usepackage{graphicx}

\begin{document}


\title{First-principles study of Co concentration and interfacial resonance states in Fe$_{1-x}$Co$_x$ magnetic tunnel junctions}


\author {J.P. Trinastic}
\author {Yan Wang}
\author {Hai-Ping Cheng}
 \affiliation {Department of Physics and Quantum Theory Project, University of Florida, Gainesville, Florida, 32611, USA}


\date{\today}

\begin{abstract}
The optimal Co concentration in Fe$_{1-x}$Co$_{x}$/MgO magnetic tunnel junctions (MTJs) that maximizes tunneling magnetoresistance (TMR) is still under investigation. We perform a first-principles transport study on MTJs using disordered electrodes modeled using the virtual crystal approximation (VCA) and ordered alloys with various MgO barrier thicknesses.  We find that 10-20$\%$ Co concentration maximizes TMR using VCA to represent disorder in the electrodes.  This TMR peak arises due to a minority d-type interfacial resonance state (IRS) that becomes filled with small Co doping, leading to a decrease in antiparallel conductance.  Calculations with ordered Fe$_{1-x}$Co$_{x}$ electrodes confirm the filling of this minority d-type IRS for small Co concentrations.  In addition, we construct a 10x10 supercell without VCA to explicitly represent disorder at the  Fe$_{1-x}$Co$_x$/MgO interface, which demonstrates a quenching of the minority-d IRS and significant reduction in available states at the Fermi level that agrees with VCA calculations.  These results explain recent experimental findings and provide implications for the impact of IRS on conductance and TMR in Fe$_{1-x}$Co$_x$/MgO tunnel junctions.
\end{abstract}


\maketitle


\begin{center}\section{Introduction}\end{center}
\label{Introduction}

Magnetic tunnel junctions (MTJs) have been studied extensively since Julliere's theoretical model proposed a differential magnetoresistance when the magnetic moments of two ferromagnetic electrodes sandwiching an insulating barrier are switched between parallel and antiparallel states~\cite{Julliere1975225}.  Industrial applications of MTJs now include memory read heads for disk drives, magnetic sensors, and magnetoresistive random access memory (MRAM)~\cite{4818515}. The difference in resistance arises from the half-metallic features of the electrodes in which only the majority or minority carriers have nonzero density of states at the Fermi energy.  In addition to half-metallicity, depending on the choice of insulating barrier, electrons with different wavefunction symmetries will decay at different rates inside the barrier, which can lead to substantial changes in magnetoresistance even for electrodes that are not completely half-metallic~\cite{Butler2008100}. These factors, known as spin-dependent tunneling, will lead to different conductances when the two electrodes have their spins oriented parallel or antiparallel to one another ($G_P$ and $G_{AP}$, respectively).  The tunneling magnetoresistance (TMR) is then defined as
\begin{equation}
TMR = \frac{(G_P - G_{AP})}{G_{AP}}
\end{equation}
Early work in tunnel junctions has found moderate success using amorphous Al$_2$O$_3$ insulating barriers sandwiched between two ferromagnetic electrodes~\cite{PhysRevLett.74.3273}.  However, recent experimental studies report TMR between 200-375 percent using a crystalline MgO barrier epitaxially matched to bcc Fe electrodes~\cite{PhysRevLett.108.176602,lee:212507,teixeira:262506}.  MgO barriers lead to higher TMR compared to  Al$_2$O$_3$ due to the MgO barrier filtering of particular wavefunction symmetries~\cite{PhysRevB.63.054416}.  Due to the cubic symmetry of the MgO lattice, electron wavefunctions with $\Delta_{1}$ symmetry (s, p$_z$, d$_{z^2}$) decay much more slowly than those with $\Delta_{2}$ (d$_{x^2-y^2}$) and $\Delta_{5}$ (p$_{xz}$, p$_{yz}$, d$_{xz}$, d$_{yz}$) symmetries within the barrier.  bcc Fe has majority $\Delta_{1}$ and $\Delta_{5}$  bands crossing the Fermi energy along the (001) direction, and the $\Delta_{1}$ states dominate parallel transmission due to their slower decay.  In the minority channel, only $\Delta_{2}$ and $\Delta_{5}$ states cross the Fermi level, which decay much faster than $\Delta_{1}$ states and leads to a small antiparallel transmission and hence a large TMR.  In this way, symmetry filtering can mimic the behavior of half-metals and are equally important to tunnel junction physics.  Subsequent first-principles studies demonstrate even higher theoretical TMR values for bcc Co/MgO/Co and ordered bcc Fe$_{0.50}$Co$_{0.50}$ junctions, which has been attributed to the fact that no $\Delta_{2}$ and $\Delta_{5}$ minority states are available for antiparallel transmission along the (001) direction when using Co leads~\cite{PhysRevB.70.172407,wang:172501}. Inspired by these findings, experimental research and industry have focused on FeCo alloys as the best potential electrode for maximizing TMR~\cite{PhysRevLett.108.176602}.

It is important, however, to take note of the differences between the previous first-principles studies and the current methodology used to fabricate FeCo MTJ electrodes.  An early study by Zhang et al~\cite{PhysRevB.70.172407} only assesses TMR in junctions with ordered Fe$_{0.50}$Co$_{0.50}$ electrodes in which the Co atom always resides at the interface between the electrode and barrier.  However, most experimental MTJs likely use disordered alloy electrodes due to annealing between 250-525 C that will lead to a mix of Fe and Co atoms at the MgO interface~\cite{2006ApPhL..89d2505Y,4836745,yuasa:222508}.  Chemical bonding at the interface and interfacial resonance states (IRS) that couple to Bloch states both significantly impact TMR, and therefore it is crucial to accurately represent the electrode surface to understand TMR results~\cite{0953-8984-15-41-R01}.  Furthermore, recent experiments suggest a nonmonotonic dependence of TMR on Co composition in Fe$_{1-x}$Co$_x$ electrodes, with the peak TMR value occuring at x=0.25~\cite{PhysRevLett.108.176602,lee:212507} as well as pure Fe and Fe$_{0.50}$Co$_{0.50}$ electrodes demonstrating similar TMR.  These recent experimental findings do not match current knowledge from first-principles transport studies, and more research examining transport using disordered alloy electrodes would help inform development of MTJs using the most effective material compositions.

A possible explanation for the TMR dependence on Co concentration is a minority IRS with $\Delta_{1}$ symmetry recently discovered at the interface between Fe$_{1-x}$Co$_x$ disordered electrodes and MgO using angle-resolved photoemission spectroscopy (ARPES)~\cite{PhysRevLett.108.176602}.  According to the ARPES results, this IRS is located above the Fermi level for pure bcc Fe electrodes, but crosses the Fermi energy at a Co concentration between 25-50 percent.  This increase in minority density of states (DOS) is postulated to create an antiparallel channel that increases antiparallel conductance and decreases TMR for higher Co concentrations.  Previous research has identified multiple bcc Fe surface states, including a mixed d$_{z^2}$/d$_{xz+yz}$ minority surface state near the Fermi level and a d$_{z^2}$ minority state around 1.8 eV above the Fermi level~\cite{PhysRevB.68.045422,PhysRevB.80.184430,ISI:000223717600046}, however it is unclear whether Bonell et al~\cite{PhysRevLett.108.176602} has identified the d$_{z^2}$ state (which has $\Delta_{1}$ symmetry) or a previously unidentified IRS.  Further insight from first-principles studies of Fe$_{1-x}$Co$_x$/MgO IRS and transport would give valuable insight into the mechanisms behind the Co concentration dependence of TMR seen experimentally.  

Several theoretical methods exist to treat disorder using density functional theory.  The virtual crystal approximation (VCA)~\cite{nordheim1931elektronentheorie,PhysRevB.61.7877} replaces all atoms within the disordered system with virtual atoms that are described by an interpolation of the relevant atomic pseudopotentials and atomic number.  VCA is computationally efficient and should theroetically work well whenever inhomogeneity can be represented through averaging.  Another method, the coherent potential approximation (CPA)~\cite{soven1967coherent}, replaces the inhomogeneous potential of a disordered material with an effective potential that is created self-consistently.  CPA has been implemented previously using a muffin-tin approximation in which the effective potential for each atom is represented using non-overlapping spheres.  Green's function expansions are then used to treat interatomic scattering~\cite{PhysRevB.40.12164,PhysRevB.31.3260,faulkner201050}.  Other options include configurational averaging of many large supercells as well as linear response theory applications~\cite{PhysRevLett.72.4001}.  Both of these options are very computationally expensive and are therefore of limited use.

In this paper, we use first-principles calculations to calculate the ballistic transport and TMR for Fe$_{1-x}$Co$_x$/MgO tunnel junctions with disordered alloy electrodes modeled using VCA.  VCA as a method of treating disorder has the important advantages of being technically simple, adding no computational costs above that required for the comparable ordered system, and is implementable in plane-wave codes.  However, since every atom is identical, this approximation does not introduce any local inhomogeneities that may be important.  Therefore, we compare these results to transport/TMR in ordered  Fe$_{0.75}$Co$_{0.25}$/MgO and  Fe$_{0.50}$Co$_{0.50}$/MgO junctions and examine differences in transport results.  Furthermore, we analyze the partial density of states (PDOS) at the interface between the electrode and MgO and compare the results to the PDOS within CPA as well as using an explicit-disordered model using a larger supercell that explicitly treats disorder within the electrode to understand how the IRS affects transport in Fe$_{1-x}$Co$_x$/MgO systems.  Since we can create VCA pseudopotentials for any relative percentage of Fe or Co, the VCA calculations provide important results about how IRS shift with increasing Co concentration and their impact on TMR.  The ordered electrode results will then provide important complementary information regarding whether VCA leaves out important contributions to transport.  By comparing VCA and ordered electrode results and determining which better matches experimental results, we can understand what aspects of disorder are most important to accurately model transport in Fe$_{1-x}$Co$_x$/MgO tunnel junctions.
\newline

\begin{center}\textbf{II. METHODS}\end{center}
\label{Methods}

\begin{center}\textbf{A. Exchange-Correlation Functional and Band Structure}\end{center}

We use the plane-wave basis set implemented in Quantum Espresso~\cite{0953-8984-21-39-395502} to solve the Kohn-Sham equations. A 10x10x10 and 10x10x1  Monkhorst-Pack~\cite{PhysRevB.13.5188} k-point mesh have been used for the Brillouin zone intergration of the leads and scattering region, respectively, with energy cutoffs for the wavefunction and charge density of 40 and 480 Ry, respectively.

All self-consistent calculations of the leads and scattering regions have been performed using the Perdew-Zunger local density approximation (PZ-LDA) of density functional theory (DFT).  We have chosen to fix the bcc Fe and Co lattice constants to experimental values of 2.87 $\textup{\AA}$ and 2.82 $\textup{\AA}$ to compare with previous first-principles calculations using the same geometry ~\cite{PhysRevB.70.172407,bose2010tailoring,miura2012first}, and we have found that in this case the Perdew-Burke-Ernzerhof generalized gradient approximation (PBE-GGA) predicts incorrect majority band structure along the $\Gamma$-H high symmetry (001) direction.  As seen in Figure 1(a) (red line), LDA results predict the $\Delta_5$ bands to cross the Fermi level near H, however GGA predicts that these bands never cross (Figure 1(b)).  These bands have significant impact on the $\Gamma$-point transmission in antiparallel alignment, as has been discussed previously~\cite{PhysRevB.70.172407} and will be discussed in the Results (Sec. III.A), and therefore it is important to capture them accurately.  Identical calculations using the projector-augmented wave (PAW) method with GGA predict the same band structure as LDA.  In addition, we find that if we relax the structure using PBE-GGA to its optimized lattice constant (2.85 $\textup{\AA}$), its band structure converges to that matching LDA results.  This suggests that the band structure using PBE-GGA is more sensitive to the value of the lattice constant compared to PZ-LDA.  Therefore, we conclude that LDA will provide more accurate results while fixing the lattice constants to experimental values.

\begin{center}\textbf{B. Virtual Crystal Approximation}\end{center}

We have used the virtual crystal approximation (VCA) to represent disordered Fe$_{1-x}$Co$_x$ alloys as leads in transport calculations.  VCA analyses are computationally efficient and have been used to describe the structural, electronic, and magnetic properties of various materials that involve transition metals~\cite{wang:123906,PhysRevB.30.259,Söderlind19922037}.  Specifically, the approximation has been found to provide accurate electronic and magnetic properties for FeCo alloys~\cite{Söderlind19922037,wang:123906}.  The approximation replaces single Fe or Co atoms in the supercell with one atom type at each lattice point described by an alloy pseudopotential and alloy charge number given as:\newline
\begin{equation}
V = (1-x) V_{Fe} + xV_{Co}
\end{equation}
\begin{equation}
Z = (1-x)Z_{Fe} + xZ_{Co}
\end{equation}
Seven virtual pseudopotentials have been generated to describe disordered Fe$_{1-x}$Co$_x$ electrodes with Co concentration from x=0.1 to x=0.7 in 0.1 intervals using the virtual.x utility in Quantum Espresso~\cite{0953-8984-21-39-395502}.  Fe$_{0.20}$Co$_{0.80}$ and Fe$_{0.10}$Co$_{0.90}$ disordered alloys have not been generated because they have been found to relax to the hcp structure and are not relevant to bcc MTJs~\cite{PhysRevLett.108.176602}.

To confirm the accuracy of VCA for the purposes of this paper, we have plotted the magnetic moment per atom of the bulk bcc Fe$_{1-x}$Co$_x$ alloys found from self-consistent calculations (Figure 2(a)) as well as the majority and minority density of states (DOS) as a function of selected Co concentrations (Figure 2(b)).  The Fe and Co structures have been fixed at the experimental lattice constants of 2.87 $\textup{\AA}$ and 2.82 $\textup{\AA}$, respectively, and lattice constants for the alloys with varying Co concentration have been linearly interpolated.  As shown in the graph, the magnetic moment per atom closely resembles the results found experimentally~\cite{1993ferr.book.....B} with a peak around x=0.25, however the DFT calculations slightly underestimate values near the peak at x=0.25-0.30 and overestimate values for higher Co concentrations.  A similar magnetic moment calculation using the coherent potential approximation (CPA) within the Layer Korringa-Kohn-Rostoker (LKKR) implementation of the local spin-density approximation of DFT reveals a similar trend as VCA but with a smaller peak magnetic moment magnitude~\cite{PhysRevB.59.5470} (Figure 1(a)).  In addition, the DOS calculations generally match previous theoretical calculations using the coherent potential approximation (CPA)~\cite{0305-4608-18-8-017}.  In particular, the majority peak around 1 eV below the Fermi level for pure Fe shifts farther below for low Co concentration before settling in a region around 2 eV below the Fermi level at higher Co concentrations.  The minority double peak around 2 eV above the Fermi level for pure Fe moves consistently closer to the Fermi level for increasing Co concentration, and the peak closest to the Fermi level gradually increases in height.  Both of these results support the use of the VCA approximation in the following transport calculations.

\begin{center}\textbf{C. Scattering Region Supercells}\end{center}

Two supercells have been constructed to compare the TMR of MTJs composed of disordered and ordered FeCo alloys.  The first supercell is the minimal size necessary to represent the Fe$_{1-x}$Co$_x$/MgO MTJ with disordered electrodes using the VCA approximation.  In this case, each lead includes two atoms per supercell at bcc basis positions.  The scattering region contains eight Fe$_{1-x}$Co$_x$ monolayers (ML) acting as buffers on each side of the MgO barrier (Figure 3(a)).  The  Fe$_{1-x}$Co$_x$ atom at the interface has been placed above the O atom as has been shown previously~\cite{PhysRevB.63.054416}.  To explore the effects of barrier thickness on TMR for each of the alloys, supercells with 6ML and 8ML MgO barriers have been constructed.  The MgO lattice constant has been fixed to be $\sqrt{2}$ factor larger than the electrode lattice to represent an epitaxial matching between the electrode and barrier.  The interlayer distance inside the MgO barrier is 2.11 $\textup{\AA}$.\newline
\indent In order to compare VCA and ordered electrodes, the second supercell is enlarged to construct ordered Fe$_{1-x}$Co$_x$ electrodes.  In this case, leads are twice as long in the z-direction (parallel to the axis of the junction) and twice as long in the x and y directions to construct Fe$_{0.75}$Co$_{0.25}$ and  Fe$_{0.50}$Co$_{0.50}$ ordered electrodes.  The scattering region (Figure 3(b)) is similarly enlarged.  Co atoms have been placed in an ordered fashion such that one Co atom is located at the FeCo/MgO interface for  Fe$_{0.75}$Co$_{0.25}$ and two Co atoms are located at the interface for ordered Fe$_{0.50}$Co$_{0.50}$.  This is a different but more realistic structure than previous first-principles calculations of ordered Fe$_{0.50}$Co$_{0.50}$~\cite{PhysRevB.70.172407}, in which the smaller unit cell was used and only Co atoms were located at the MgO interface.  The construction of the present unit cell and Fe/MgO interface better represents experimental conditions with a mix of Fe and Co at the interface~\cite{moriyama:222503,shikada:07C303}.\newline
\indent To determine the optimal distance between the transition metal and oxygen atom at the tunnel junction interface, self-consistent energy calculations have been done for various Fe/Co-O distances (from 2 to 2.3 $\textup{\AA}$ in 0.05 $\textup{\AA}$ intervals).  A 2.20 $\textup{\AA}$ Fe/Co-O distance provides the lowest energy structure, similar to first-principles results for similar systems~\cite{burton:142507,PhysRevB.70.172407}.\newline
\indent Tunneling conductance for parallel and antiparallel electrode magnetization configurations has been calculated using the PWCOND~\cite{PhysRevB.70.045417} code within Quantum Espresso.  This code solves for transmission coefficients by matching Bloch state wavefunctions in the bulk leads and scattering regions and calculates the total conductance through the junction using the Landauer-Buttiker formula:
\begin{equation}
G = \frac{e^2}{h}\sum_{\textbf{k}_{||}}^n T(k)
\end{equation}
where the sum is over the components of the wave vectors parallel to the junction interface ($\textbf{k}$$_{||}$).  A separate conductance is calculated for each spin channel: 1) parallel majority ($G_P^{maj}$), 2) parallel minority ($G_P^{min}$), and 3) antiparallel ($G_{AP}$) (majority and minority antiparallel channels are equal due to the symmetry of the junction).  The TMR can then be calculated using equation (1), where $G_P = G_P^{maj} + G_P^{min}$.

\begin{center}\textbf{III. RESULTS AND DISCUSSION}\end{center}
\label{results}

\begin{center}\textbf{A. VCA Transport}\end{center}

In Figure 4(a-c, blue and green lines), we show the calculated conductances for each spin channel using VCA electrodes as a function of Co concentration for different thicknesses (6 and 8 ML).   For each channel, conductance is calculated as an integral summation of the transmission probabilities over the 2D Brillouin Zone (2DBZ).  Majority parallel and antiparallel conductance values converge using a 100x100 k-point grid.  A 400x400 k-point grid is necessary for minority parallel conductance convergence due to the existence of sharply peaked interfacial resonance states (IRS)  that influence transmission in this channel~\cite{PhysRevB.70.172407,wang:172501,PhysRevB.82.054405}.

As seen in Figure 4(a), the majority parallel conductance $G_P^{maj}$ decreases slowly with increasing Co concentration as expected due to the majority density of states decreasing at the Fermi level as Co is added (Figure 2(b)).  The minority parallel conductance $G_P^{min}$ decreases sharply at x=0.2, and then increases slightly to a steady value from x=0.4 to x=0.7 (Figure 4(b)).  The value of $G_P^{min}$ across all Co concentrations, however, is orders of magnitude less than $G_P$, and therefore does not play a significant role in TMR.  The $G_P^{min}$ behavior is likely due to an IRS, which we later confirm using projected density of states (PDOS) calculations, in which electronic states are projected onto orthogonalized atomic orbitals on each atom to identify wavefunction symmetries.  In addition, PDOS analysis can identify IRS that are present on interface Fe/Co atoms but disappear in the bulk.  Previous studies have shown that IRS strongly influence conductance through an Fe/MgO junction~\cite{ ISI:000223717600046}.  IRS arise when the two-dimensional dispersion of states at the Fe/MgO interface cross bands in the bulk electrode~\cite{PhysRevLett.75.2960}.  IRS on both sides of the barrier can couple to one another, thus generating the resonance effect and increasing conductance, but the coupling strength decreases exponentially with barrier thickness~\cite{Tiusan2006112}.

In contrast, the antiparallel conductance behavior, $G_{AP}$, has a greater impact on TMR, decreasing from x = 0 to x = 0.2 for the 6ML junction and from x = 0 to x = 0.1 in the 8ML junction (Figure 4(c)).  After reaching a minimum value for these low Co concentrations, $G_{AP}$ continues to rise up to x=0.7, after which it decreases again for pure Co electrodes.  This decrease in $G_{AP}$ for small Co concentration diminishes as the barrier thickness increases, again suggesting the role of IRS in the conductance behavior.  We confirm this by plotting the 2D transmission probability as a function of $\textbf{k}$$_{||}$ in Figure 5 for each MgO barrier thickness.  For the 6ML barrier, IRS can be clearly seen as sharply peaked, high-probability transmission at points close to k$_x$ = 0 and k$_y$ = 0 for x = 0 and x = 0.1.  At x = 0.2, these IRS completely disappear, leading to a drop in $G_{AP}$ for the 6ML barrier.  For x $>$ 0.25, transmission probabilities become higher in a circular region around $\textbf{k}$$_{||}$ = 0 that grows larger with increasing Co that leads to an increase in $G_{AP}$.  The IRS hot spots almost completely disappear for the 8ML barrier, leading to transmission being dominated by the regions close to $\textbf{k}$$_{||}$ = 0.  Based on these conductance values, the tunneling magnetoresistance (TMR), as calculated by Equation (1), is shown as a function of Co concentration in Figure 4(d) for tunnel junctions with 6 and 8 MgO layers.  TMR is maximized with a 20 percent Co concentration in the 6ML junction and a 10 percent Co concentration in the 8ML junction.  It is clear from Figure 4(d) that for electrodes with x $<$ 0.5, there is a strong dependence of TMR on barrier thickness, however for x=0.5 and greater, TMR converges to fairly stable values for the 6ML and 8ML barriers.  This result matches previous experimental findings that have shown an increase in TMR with barrier thickness using Fe electrodes up to about 2 nm MgO thickness~\cite{yuasa2004giant}.  In addtion, recent experiment has not shown this strong thickness effect when using CoFeB electrodes that have been annealed to the same temperature~\cite{hayakawa2005dependence}.  Our results indicate that this thickness effect is strongest with pure Fe electrodes due to the IRS peaks that significantly increase $G_{AP}$ for Fe-heavy electrodes in thinner MgO barriers, thus leading to large TMR increase when the transmission hot spots disappear for low Co concentrations.  Once the IRS do not play a large role, for x $>$ 0.5, TMR is similar for both barrier thicknesses.

As described above, the conductance behavior suggests an important role of IRS as already posited in previous studies of Fe/MgO junctions.  To explain the majority and minority conductance behaviors in detail, we calculate the partial density of states (PDOS) for the Fe$_{1-x}$Co$_x$/MgO  interfacial layer to determine the evolution of the IRS as Co concentration (x) increases.  Figure 6(a) shows the minority s-PDOS of the Fe$_{1-x}$Co$_x$ interface layer using VCA for various Co concentrations (not all Co concentration values are shown for the sake of graph clarity, however samples are shown to demonstrate the overall trend).  The peak around 0.25 eV above the Fermi level is clearly an IRS of pure Fe (x=0.0, red line), as it disappears in the bulk electrode two monolayers from the interface (dotted line).  As Co concentration increases in the electrode, the IRS shifts closer to the Fermi Energy.  The peak is just above the Fermi level for x = 0.25 and below for x = 0.50, indicating it will be filled around x=0.30.  An IRS at this position matches recent ARPES result indicating an IRS that becomes filled between x = 0.25 and x = 0.50~\cite{PhysRevLett.108.176602}.  However, if this IRS played a significant role in transport, we would expect to see transmission hot spots for $G_{AP}$, around x = 0.30, which we do not see in our results (Figure 5).  This suggests that this s-type IRS does not play a significant role in transport, but rather for x $>$ 0.3, Co minority states begin to be accessible at the Fermi level (see DOS in Figure 2(b)),  which opens minority channels and explains why $G_P^{min}$ and $G_{AP}$ do not decrease further.  To the best of our knowledge, this is the only first-principles confirmation of this IRS, which has recently been found experimentally at the Fermi level at a Co concentration between 25 and 50 percent~\cite{PhysRevLett.108.176602}.

A similar analysis can be done to understand the significant drop in $G_P^{min}$ and $G_{AP}$ when comparing pure Fe electrodes to electrodes with small Co concentrations within the VCA calculations.   As seen in Figure 6(b), the minority d-projected PDOS demonstrate a mixed d$_{z^2}$/d$_{xz+yz}$ IRS about 0.08 eV above the Fermi level for a pure Fe electrode as well as a d$_{z^2}$ state around 1.8 eV above the Fermi level.   These states have been previously identified experimentally as well on Fe surfaces~\cite{PhysRevB.68.045422}.  The pure d$_{z^2}$ surface state is unoccupied for all Co concentrations and never comes close to the Fermi level; therefore, we will not discuss it further here as it does not play a role in transport.  However, the mixed  d$_{z^2}$/d$_{xz+yz}$ state is partially filled for pure Fe, which leads to greater $G_{AP}$, explaining the higher $G_{AP}$ when using pure Fe electrodes as well as the hot spots identified on the transmission probability graphs (Figure 5).  However, as Co concentration increases, this IRS broadens and becomes partially filled, leaving no states to allow for electron transport and significantly reducing $G_{AP}$ for the 6ML barrier as seen in Figure 4(c) for small Co concentrations.  At x=0.25 and greater, new transmission channels in the circular region close to $\textbf{k}$$_{||}$ = 0 due to increased Co concentration prevents further  G$_{AP}$ decrease and results in a TMR peak at x=0.20 (Figure 4(d), blue line).  As barrier thickness increases, transmission at k-points far from $\textbf{k}$$_{||}$ = 0 decay more quickly than those close to $\textbf{k}$$_{||}$ = 0~\cite{Butler2008100}.  Therefore, for the 8ML barrier, the IRS have significantly decayed and transmission is dominated by regions close to $\textbf{k}$$_{||}$ = 0 even for smaller Co concentrations.  This leads to the $G_{AP}$ minimum and TMR maximum occurring at x=0.1 for the 8ML barrier (Figure 4(d), green line).  These results indicate that the minority d$_{z^2}$/d$_{xz+yz}$ state is the most significant factor in determining TMR for Fe$_{1-x}$Co$_x$/MgO electrodes, as the significant drop in $G_{AP}$ when this IRS becomes filled leads to the maximum TMR at x=0.2 and x=0.1 for the 6ML and 8ML barriers, respectively.

The TMR pattern  using the 6ML MgO barrier are extremely close to recent experimental findings that report a maximum TMR value at 25 percent Co concentration using a 1.5-2.5 nm MgO barrier~\cite{PhysRevLett.108.176602,lee:212507} and a similar TMR for pure Fe and Fe$_{0.5}$Co$_{0.5}$ electrodes~\cite{yuasa:222508,PhysRevLett.108.176602,lee:212507}.  The large TMR using pure Fe electrodes with an 8ML MgO barrier does not match experimental data, however this is likely due to the fact that the nominal MgO thickness used in experiment is often different than the effective or actual thickness~\cite{PhysRevLett.97.087210}.  In addition, FeO layers or oxygen vacancies are known to develop at the interface in experiments and reduce TMR that we did not model in this study~\cite{PhysRevLett.87.076102,PhysRevB.68.092402}.  It is important to note that oxygen diffusion mechanisms in actual physical systems not accounted for in the present study can also enhance the interface coupling across the barrier~\cite{0953-8984-19-16-165201}.  This would sustain the effect of IRS for thicker barriers, in which case our results for the TMR peak at 20 percent Co for the 6ML junction may better reflect experimental situations and are in good agreement with previous experimental results~\cite{PhysRevLett.108.176602,lee:212507}.  Future theroetical and experimental studies are important to understand how various levels of Co doping in the electrode affect interfacial effects such as FeO layer formation.  In addition, our study reports TMR for more Co concentration intervals compared to experiment, and our results suggest that even less Co (20 vs 25 percent) is required to optimize TMR.  It will be important for future experiments to test a finer interval of Co concentrations to confirm this finding.

It is important to note that our TMR results for pure bcc Fe and Co electrodes differ from previous first-principle results using LKKR that predicts the Co/MgO(8 ML)/Co junction to have a larger TMR compared to the Fe/MgO(8 ML)/Fe junction~\cite{PhysRevB.70.172407}.  This finding has been explained as follows assuming the long barrier limit in which transmsision at $\textbf{k}$$_{||}$ = 0 dominates.  For bcc Co electrodes, only a $\Delta_1$ majority band crosses the Fermi level along the (001) direction, however no minority $\Delta_1$ band crosses the Fermi level.  Thus, in antiparallel alignment, there is total reflection at $\textbf{k}$$_{||}$ = 0 because there are no minority $\Delta_1$ states available to couple to the majority $\Delta_1$ state.  On the other hand, bcc Fe has both majority and minority $\Delta_5$ states crossing the Fermi level that lead to a small but finite transmission probability at $\textbf{k}$$_{||}$ = 0 compared to bcc Co.  As shown in Figure 7, our results confirm the significantly reduced transmission at $\textbf{k}$$_{||}$ = 0 for  Co compared to Fe electrodes.  However, our results for all barrier thicknesses demonstrate that $\textbf{k}$$_{||}$ $\neq$ 0 significantly contribute to $G_{AP}$, even when IRS do not play a large role as in the 8ML barrier (Figure 5).  Because of this, the areas of higher transmission when using Co electrodes lead to a finite $G_{AP}$ comparable to Fe that decreases its TMR.  Previous first-principles studies have also found a higher TMR using bcc Fe compared to bcc Co electrodes~\cite{bose2010tailoring}, confirming that the $\textbf{k}$$_{||}$ $\neq$ 0 are important.  Previous theoretical and experimental work has found higher TMR when adding bcc Co interlayers between the bcc Fe electrode and MgO barrier~\cite{yuasa:222508,wang:172501}, however our present results suggest a pure bcc Co electrode will not necessarily improve TMR.  

\begin{center}\textbf{B. Ordered FeCo Electrode Transport Calculations}\end{center}

We next calculate the conductance for each spin channel and TMR for ordered Fe$_{0.75}$Co$_{0.25}$ and  Fe$_{0.50}$Co$_{0.50}$ electrodes using the larger unit cell (as described in Sec II.C).  Each conductance is calculated as an integral summation of the transmission probabilities over the 2D Brillouin Zone (2DBZ).

Figure 4(a-d, red and black lines) shows the conductance per spin channel and TMR for the ordered electrode tunnel junctions for each barrier thickness.  Using ordered electrodes, $G_P^{maj}$ decreases slowly with increased Co concentration (Figure 4(a)) across all barrier thicknesses and is identical to VCA results.  Similar to VCA calculations, $G_P^{min}$ decreases for Fe$_{0.75}$Co$_{0.25}$ and then increases for the Fe$_{0.50}$Co$_{0.50}$ ordered electrode in the 6ML and 8ML junctions, however the conductance in both cases is larger compared to the VCA results.  Similar to the VCA results, $G_P^{min}$ is orders of magnitude smaller than $G_P^{maj}$ and therefore plays little role in TMR.  The antiparallel conductance $G_{AP}$ shows a similar pattern to the VCA results, however the magnitude is slightly larger for both Fe$_{0.75}$Co$_{0.25}$ and Fe$_{0.50}$Co$_{0.50}$ in the 6ML and 8ML cases compared to VCA.  Despite this difference, the TMR pattern is the same, with Fe$_{0.75}$Co$_{0.25}$ leading to the maximum TMR in the 6ML case and pure Fe in the 8ML cases.  

We next examine IRS of the ordered Fe$_{1-x}$Co$_x$/MgO interfaces and compare them to the VCA IRS described above to understand if they may lead to the differences in conductance.  As plotted in Figure 8(a), a similar behavior for the minority s-type IRS is seen in the ordered electrode-MgO interface, however the IRS for the  Fe$_{0.75}$Co$_{0.25}$ electrode is broadened compared to the VCA cases and partially filled.  This may open a minority channel that will increase $G_P^{min}$ and $G_{AP}$ comapred to the VCA case, possibly explaining the differences in conductance between VCA and ordered electrodes, however our VCA results indicate that this IRS did not play a significant role in transport.  The minority d$_{z^2}$/d$_{xz+yz}$ IRS broadens and shifts below the Fermi level for Fe$_{0.75}$Co$_{0.25}$ and Fe$_{0.50}$Co$_{0.50}$ in a similar way as the VCA PDOS, leading to the initial decrease in $G_{AP}$ and maximum TMR for the Fe$_{0.75}$Co$_{0.25}$ electrode in the 6ML case.  In general, the ordered transport results confirm the major factors influencing transport discussed in the VCA case.

Beyond IRS, several other possibilities could explain the small increase in $G_P^{min}$ and $G_{AP}$ compared to the VCA calculations.  In the VCA case, each atom is identical; therefore, as more Co is added, the band structure simply shifts in the direction of the Co band structure.  This creates a drastic change in the availability of Bloch states as soon as a band shifts below the Fermi level, as the $\textbf{k}$$_{||}$ = 0 results shown in Figure 7 exemplify.  However, in the ordered transport cases, Fe bands that cross the Fermi level are still available for transport even with significant Co doping.  This  may lead to a less severe decrease in the $G_P^{min}$ and $G_{AP}$ as Co concentration increases.  Another reason may relate to the existence of quantum well states that will arise due to the periodic potential in the ordered case that does not exist in the VCA case in which each lead atom is identical.  These quantum well states could lead to resonance energies that create transmission hot spots that increase conductance.  However, more detailed study beyond the scope of this paper is required to explore these possibilities.

\begin{center}\textbf{C. CPA and Explicit-Disordered Model Calculation}\end{center}

A shortcoming of VCA is its inability to take into account the role of local distortion effects on electronic structure because each atom in the VCA electrode is treated as an identical virtual atom and sees the same potential.  To examine whether this may have an effect on the minority d$_{z^2}$/d$_{xz+yz}$ IRS that is central to understanding transport as described above, we calculate the interface PDOS using both CPA within the layer KKR framework~\cite{PhysRevB.59.5470} as well as a large supercell to explicitly model disorder (explicit-disordered model).  For the latter, we construct a 10x10 supercell of the Fe$_{1-x}$Co$_x$/MgO interface for x=0.10, x=0.25, and x=0.50 using a slab containing 3 Fe$_{1-x}$Co$_x$ monolayers (300 atoms) and 2 MgO monolayers (400 atoms) (Figure 9).  A 20 $\textup{\AA}$ vacuum is used perpendicular to the interface to separate slab images.  To represent disorder in the electrode, we randomly replace 10$\%$, 25$\%$, and 50$\%$ of the Fe electrode atoms with Co for Fe$_{0.90}$Co$_{0.10}$, Fe$_{0.75}$Co$_{0.25}$ and  Fe$_{0.50}$Co$_{0.50}$ layers, respectively.  The PDOS of the interface electrode atoms are calculated.  

As shown in Figure 10(a), the minority d-projected CPA PDOS results demonstrate similar positions of the IRS for both Fe$_{0.75}$Co$_0.25$  and Fe$_{0.50}$Co$_0.50$ electrodes as that seen in the VCA and ordered cases but with significantly greater broadening.  In addition, as shown in Figure 10(b), the minority d$_{z^2}$/d$_{xz+yz}$ IRS for the explicit-disordered model demonstrates even greater broadening compared to VCA and CPA.  The IRS diminishes significantly in height for small Co doping (x=0.10) and almost completely disappears for larger doping (x=0.25 and x=0.50).  Although the VCA calculations do not demonstrate this IRS quenching, they do closely match the significant reduction in available states at the Fermi level as seen in the large supercell calculation, which is the factor that affects the transport calculations at zero bias.  Specifically, for the larger, disordered supercell, the minority-d IRS at the Fermi level decreases to 60 percent of the pure Fe state for low Co doping (x=0.10), and decreases further to about half of the pure Fe magnitude for x=0.25 and x=0.50.  This is very similar to the minority-d IRS seen in the VCA calculations (Figure 5(b)), and may suggest an even stronger effect of low Co doping in quenching the IRS and leading to a TMR increase.  Previous studies have shown the quenching of IRS in a Fe/MgO junction by using an interlayer~\cite{PhysRevB.72.140404,PhysRevB.82.054405,wang:172501}, and disorder has been shown to quench surface states in topological insulators~\cite{PhysRevB.85.201105}.  This important result provides new information about the effect of Co concentration on Fe$_{1-x}$Co$_x$/MgO IRS and supports the VCA transport results as an accurate representation of the conductance and TMR changes in Fe$_{1-x}$Co$_x$/MgO junctions at zero bias.

\begin{center}\textbf{IV. CONCLUSIONS}\end{center}
\label{Conclusions}

This study uses VCA transport calculations to provide support for a small amount (10 - 20$\%$) Co doping of  Fe$_{1-x}$Co$_x$ electrodes to maximize TMR due to the filling of the minority d$_{z^2}$/d$_{xz+yz}$ IRS for small Co concentrations.  Our results have demonstrated a peak TMR of 13000$\%$ for a Fe$_{0.80}$Co$_{0.20}$/MgO(6ML) tunnel junction and 19000$\%$ for a Fe$_{0.90}$Co$_{0.10}$/MgO(8ML)  junction, filling an important gap in the literature and closely matching recent experiment~\cite{PhysRevLett.108.176602}.   In particular, we find even less Co required (10-20$\%$) than that used in recent experiments, which suggests further study is needed using finer doping intervals to determine the optimal level.  In addition, we find that Fe$_{0.50}$Co$_{0.50}$/MgO tunnel junctions exhibit similar (6ML) or lower (8ML) TMR compared to Fe/MgO junctions, again matching recent experiment~\cite{PhysRevLett.108.176602,lee:212507}.  The barrier lengths used in this study (1.3-1.7 nm) overlap with traditional experimental MgO thicknesses (1.5-2.5 nm)~\cite{PhysRevLett.108.176602,lee:212507}.  In addition, it is known that effects of IRS important for understanding the transport behavior persist for thicker barriers in experiment, due to diffusion and non-ideal epitaxial matching~\cite{0953-8984-19-16-165201}.  Therefore, the results presented here with the thinner MgO barrier (6ML) elucidates the important effects from IRS that are relevant to experimental design.    To this end, we provide comprehensive information about all IRS relevant to transport through Fe$_{1-x}$Co$_x$/MgO junctions and identify the importance of small Co concentrations to quench the minority-d IRS and maximize TMR.  We als confirm the existence of a minority s-type IRS previously identified experimentally~\cite{PhysRevLett.108.176602}, however we find that this IRS does not signficantly affect minority conductance channels.  Rather, beginning around 30$\%$ Co doping, minority Co states cross the Fermi level and lead to the characteristic Co transmission channels in a circular region around $\textbf{k}$$_{||}$ = 0 that increase antiparallel conductance.  This leads to a maximum TMR when using 10-20$\%$ Co.  Finally, we confirm our results by demonstrating a similar evolution of the  minority d-orbital IRS when using ordered Fe$_{1-x}$Co$_x$ electrodes as well using the CPA method.  A more realistic, explicit-disordered model using a large supercell representing the Fe$_{1-x}$Co$_x$/MgO interface suggests that the minority-d IRS is quenched for x = 0.25 and greater, confirming the VCA conductance results that the IRS only plays a role for x = 0 to x = 0.2.  By using varying barrier thicknesses, we can confirm that the changes in conductance with varying Co concentration are due to IRS, as these effects diminish for increasing barrier thickness.  

These transport calculations using the VCA electrodes provide important information about both the physical mechanisms driving the TMR maximum at small Co doping as well as the important IRS affecting transport that can guide future experimental design of MgO tunnel junctions.  Future studies should consider disordered electrodes in combination with interface roughness, Fe$_{1-x}$Co$_x$/O layers, and oxygen vacancies to determine how these realistic interface effects change the TMR results presented here.


%


\begin{figure}[h]
\includegraphics[scale=0.50]{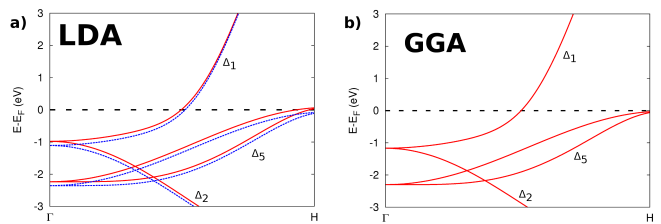}
\caption{(Color online) a) bulk Fe (red solid line) and  Fe$_{0.90}$Co$_{0.10}$ (blue dotted line) majority spin band structure along the (001) direction using LDA and experimental lattice constant of 2.87 $\textup{\AA}$; b) bulk Fe majority band structure along the (001) direction using GGA and experimental lattice constant of 2.87 $\textup{\AA}$.  The Fermi level is shown with the dotted line and orbital symmetries of each band are labeled on the graph. See text for detailed description.}
\end{figure}

\begin{figure*}[h]
\includegraphics[scale=1.00]{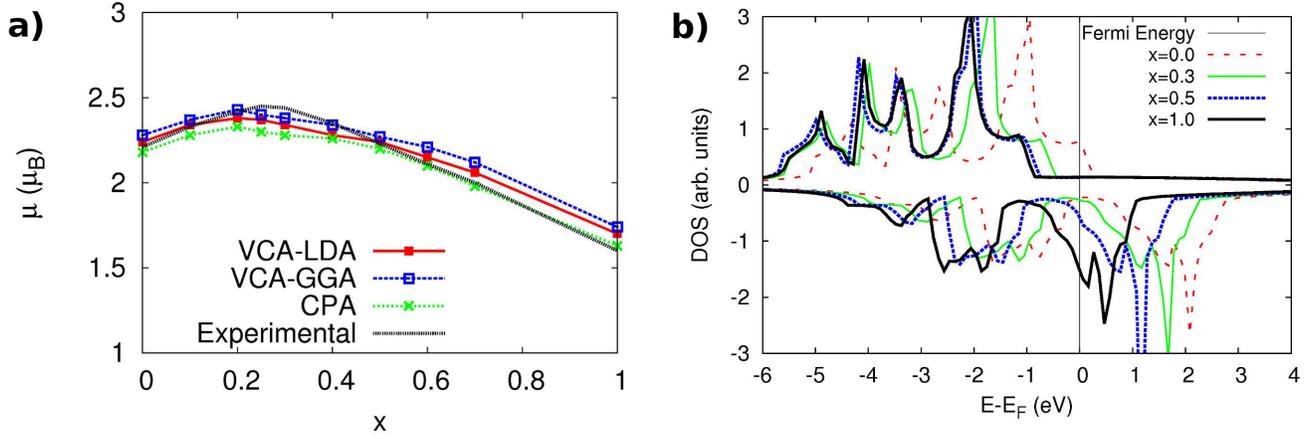}
\caption{(Color online) bcc Fe$_{1-x}$Co$_x$ bulk VCA calculations. a) VCA atomic magnetic moment as a function of Co concentration x using a one-atom VCA unit cell compared to experimental and CPA calculations (see text for VCA and CPA descriptions).  b) Density of states of  Fe$_{1-x}$Co$_x$  one-atom unit cell for x=0, 30, 50, and 100 percent Co concentrations.  Majority and minority states are above and below the horizontal line, respectively.  Not all concentrations are shown to maintain graph clarity, however the trends are visible from the sampled Co concentrations.}
\end{figure*}

\begin{figure}[h]
\includegraphics[scale=0.08]{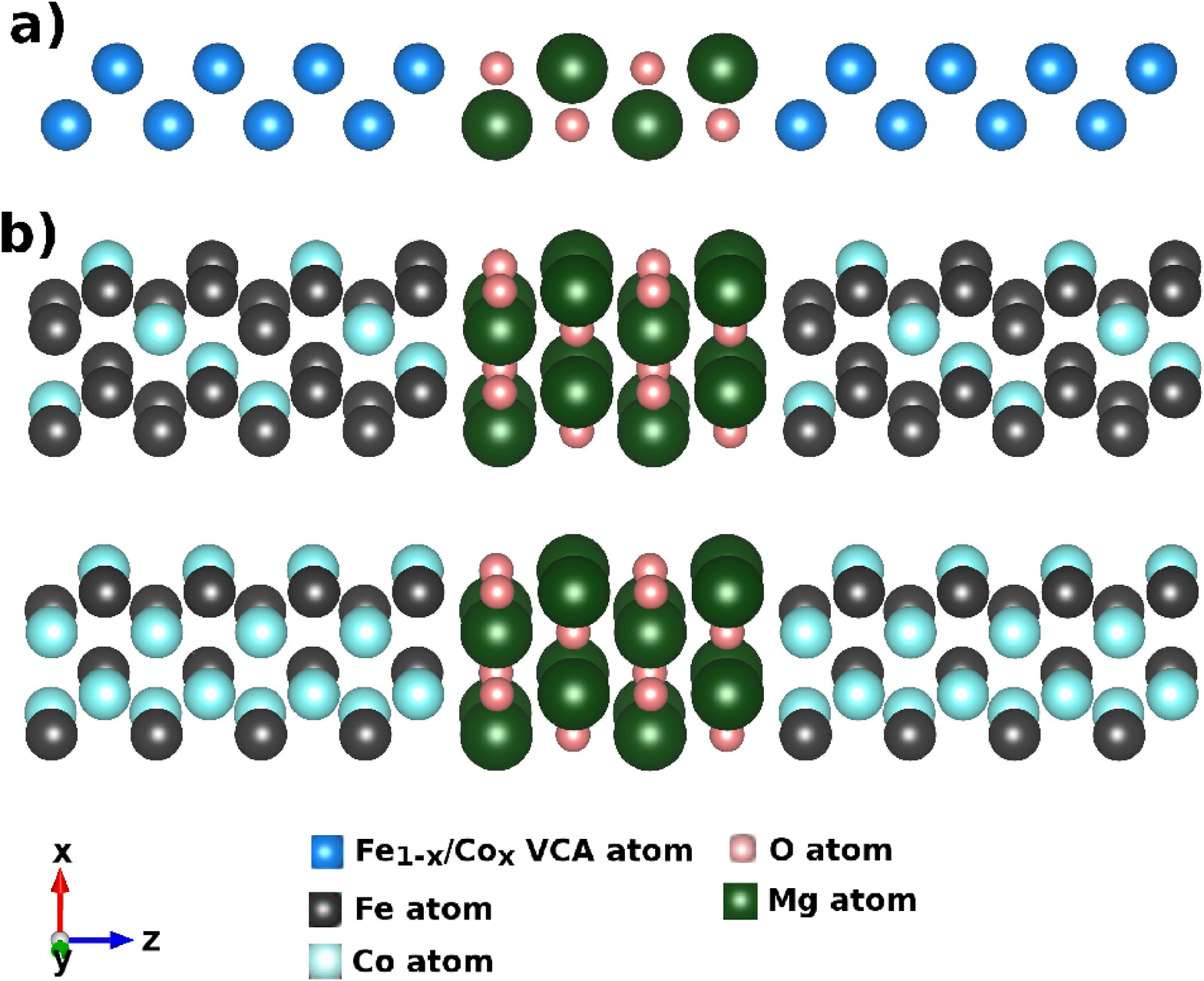}
\caption{(Color online) Fe$_{1-x}$Co$_x$/MgO scattering regions. a) Scattering region for the VCA-electrode tunnel junction.  Small light atoms (red online) and large dark atoms (green online) in the barrier region are O and Mg, respectively, and the medium-sized electrode atoms (blue online) are the Fe$_{1-x}$Co$_x$ VCA atoms.  b) Scattering region for the ordered-electrode tunnel junction.  Small light atoms (red online) and large dark atoms (green online) in the barrier region are O and Mg, respectively.  In the leads, dark atoms (gray online) are Fe and light atoms (light blue online) are Co.  In all cases, O sits closest to the electrode interface atoms as shown in experiment.}
\end{figure}

\begin{figure}[h]
\includegraphics[scale=0.75]{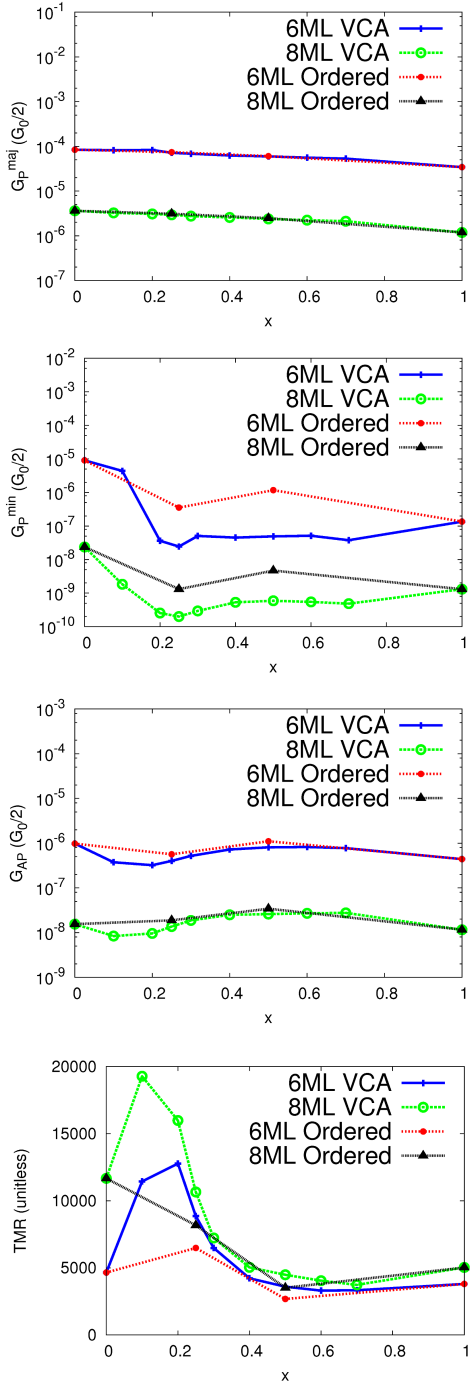}
\caption{(Color online) Conductance and tunneling magnetoresistance (TMR) as a function of Co concentration x in  Fe$_{1-x}$Co$_x$/MgO tunnel junctions with VCA and ordered electrodes using 6ML and 8ML MgO barriers. a) Majority parallel conductance G$_P^{maj}$. b) Minority parallel conductance G$_P^{min}$. c) Antiparallel conductance G$_{AP}$. d) TMR.  Since the ordered tunnel junctions are twice as long in each parallel direction, we have normalized the ordered conductance values (dividing by four) for direct comparison to the VCA electrode case.  G$_0$ is the conductance quantum.}
\end{figure}

\begin{figure*}[h]
\includegraphics[scale=0.27]{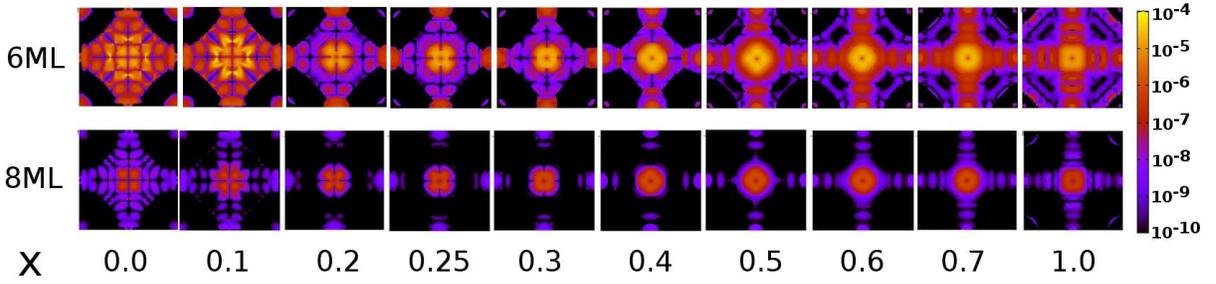}
\caption{(Color online) Transmission probability at Fermi level as a function of $\textbf{k}$$_{||}$ in the AP configuration for Fe$_{1-x}$Co$_x$/MgO(6ML) and Fe$_{1-x}$Co$_x$/MgO(8ML) junctions as labeled.  For 6ML MgO, AP conductance is largest for x=0.0 and x=0.1 in which the interface resonance states (IRS) are strongest.  For 8ML MgO, the IRS have decayed and do no play a significant role; instead, AP conductance is dominated by k-points closer to the center of the 2D Brillouin zone.}
\end{figure*}

\begin{figure*}[h]
\includegraphics[scale=1.00]{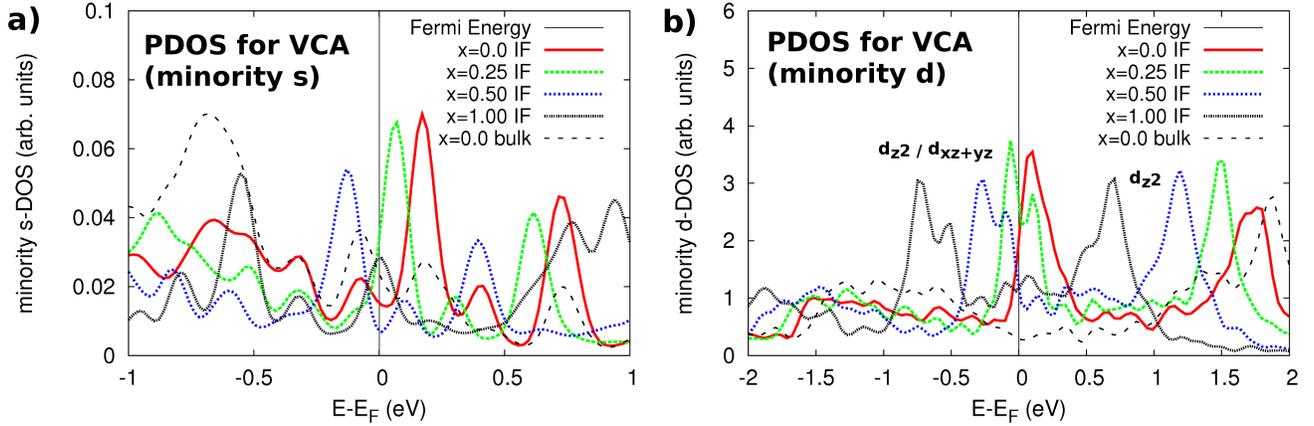}
\caption{(Color online) Partial density of states (PDOS) at the  Fe$_{1-x}$Co$_x$/MgO interface (IF) compared to Fe bulk using VCA electrodes for varying Co concentrations. a) Minority s-projected DOS.  b) Minority d-projected DOS.  The d-orbital type or mix of types of each IRS is labeled above the states in each figure.}
\end{figure*}

\begin{figure}[h]
\includegraphics[scale=0.70]{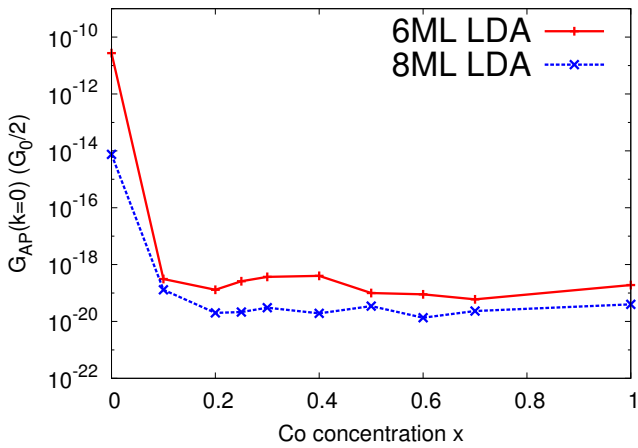}
\caption{(Color online) Transmission probability at $\textbf{k}$$_{||}$ = 0 as a function of Co concentration x in the Fe$_{1-x}$Co$_x$ VCA electrodes.  For x = 0.1 and greater, $\Delta$$_5$ bands along the (001) dierction move below the Fermi level, leading to total reflection at $\textbf{k}$$_{||}$ = 0.  See text for more detailed discussion.}
\end{figure}

\begin{figure*}[h]
\includegraphics[scale=1.00]{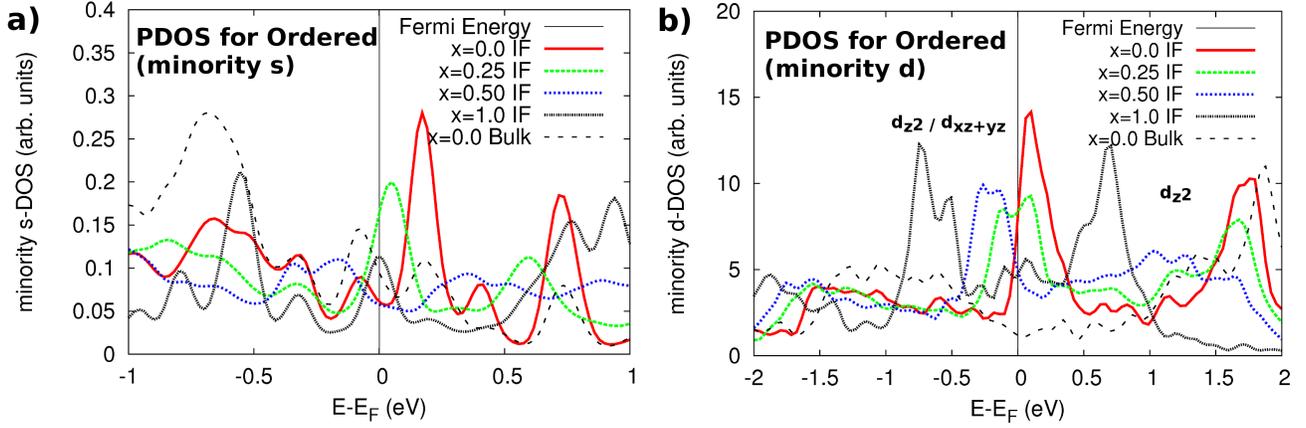}
\caption{(Color online) Partial density of states (PDOS) at the  Fe$_{1-x}$Co$_x$/MgO interface (IF) compared to Fe bulk using ordered structures for varying Co concentration. a) averaged minority s-projected DOS.  b) averaged minority d-projected DOS.  The d-orbital type or mix of types of each interface state is labeled above the states in each figure.}
\end{figure*}

\begin{figure}[h]
\includegraphics[scale=1.00]{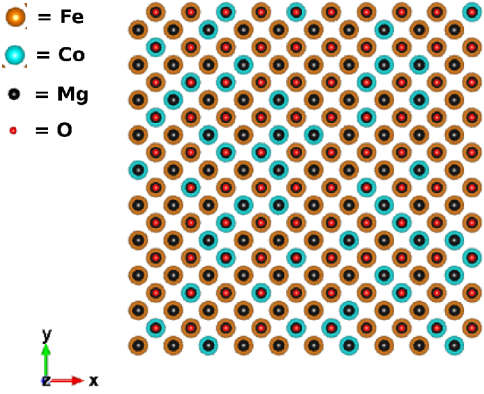}
\caption{(Color online) Example structure for the Fe$_{1-x}$Co$_x$/MgO supercell calculation representing a disordered Fe$_{1-x}$Co$_x$/MgO interface, x=0.25, referred to as the explicit-disordered model in the text.  Small light atoms (red online) and small dark atoms (black online) are O and Mg, respectively.  Dark, large atoms (orange online) are Fe and light, large atoms (light blue online) are Co.  In all cases, O sits closest to the electrode interface atoms as shown in experiment.  To generate the explicit-disordered electrode, starting with a pure Fe electrode, random number generation is used to randomly replace a quarter of the Fe atoms wth Co for x=0.25 and half of the atoms with Co for x=0.50 per layer.  Note that Co atoms seen from this perspective come from multiple layers of Fe$_{1-x}$Co$_x$, not just the interface layer.}
\end{figure}

\begin{figure*}[h]
\includegraphics[scale=1.00]{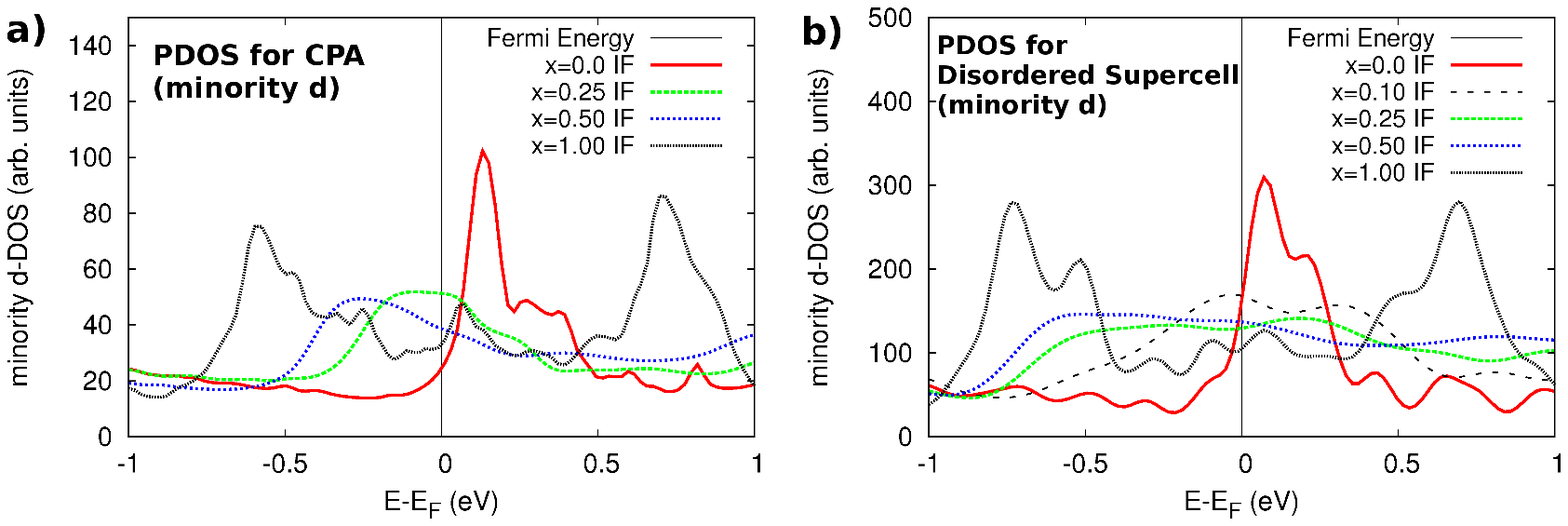}
\caption{(Color online) Partial minority-d orbital density of states at the disordered  Fe$_{1-x}$Co$_x$/MgO interface (IF) using a) the coherent potential approximation (CPA), and b) the 10x10 explicitly disordered supercell.  Only the d$_{z^2}$/d$_{xz+yz}$ interfacial resonance state is shown here as it is the most relevant for transport.}
\end{figure*}


%



\begin{acknowledgments}
\begin{center}\bf{Acknowledgments}\end{center}
	We would like to thank Dr. X.-G. Zhang for helpful discussion.  We also gratefully acknowledge financial support from the Department of Energy (Grant No. DOE/BES DE-FG02-02ER45995).  We would also like to thank The Quantum Theory Project, University of Florida High Performance Computing (UFHPC), and The National Energy Research Scientific Computing Center (NERSC) for computational resources.
\end{acknowledgments}
\bibliography{FeCoAPS101712}

\begin{thebibliography}{49}%
\makeatletter
\providecommand \@ifxundefined [1]{%
 \@ifx{#1\undefined}
}%
\providecommand \@ifnum [1]{%
 \ifnum #1\expandafter \@firstoftwo
 \else \expandafter \@secondoftwo
 \fi
}%
\providecommand \@ifx [1]{%
 \ifx #1\expandafter \@firstoftwo
 \else \expandafter \@secondoftwo
 \fi
}%
\providecommand \natexlab [1]{#1}%
\providecommand \enquote  [1]{``#1''}%
\providecommand \bibnamefont  [1]{#1}%
\providecommand \bibfnamefont [1]{#1}%
\providecommand \citenamefont [1]{#1}%
\providecommand \href@noop [0]{\@secondoftwo}%
\providecommand \href [0]{\begingroup \@sanitize@url \@href}%
\providecommand \@href[1]{\@@startlink{#1}\@@href}%
\providecommand \@@href[1]{\endgroup#1\@@endlink}%
\providecommand \@sanitize@url [0]{\catcode `\\12\catcode `\$12\catcode
  `\&12\catcode `\#12\catcode `\^12\catcode `\_12\catcode `\%12\relax}%
\providecommand \@@startlink[1]{}%
\providecommand \@@endlink[0]{}%
\providecommand \url  [0]{\begingroup\@sanitize@url \@url }%
\providecommand \@url [1]{\endgroup\@href {#1}{\urlprefix }}%
\providecommand \urlprefix  [0]{URL }%
\providecommand \Eprint [0]{\href }%
\providecommand \doibase [0]{http://dx.doi.org/}%
\providecommand \selectlanguage [0]{\@gobble}%
\providecommand \bibinfo  [0]{\@secondoftwo}%
\providecommand \bibfield  [0]{\@secondoftwo}%
\providecommand \translation [1]{[#1]}%
\providecommand \BibitemOpen [0]{}%
\providecommand \bibitemStop [0]{}%
\providecommand \bibitemNoStop [0]{.\EOS\space}%
\providecommand \EOS [0]{\spacefactor3000\relax}%
\providecommand \BibitemShut  [1]{\csname bibitem#1\endcsname}%
\let\auto@bib@innerbib\@empty
\bibitem [{\citenamefont {Julliere}(1975)}]{Julliere1975225}%
  \BibitemOpen
  \bibfield  {author} {\bibinfo {author} {\bibfnamefont {M.}~\bibnamefont
  {Julliere}},\ }\href {\doibase 10.1016/0375-9601(75)90174-7} {\bibfield
  {journal} {\bibinfo  {journal} {Physics Letters A}\ }\textbf {\bibinfo
  {volume} {54}},\ \bibinfo {pages} {225 } (\bibinfo {year}
  {1975})}\BibitemShut {NoStop}%
\bibitem [{\citenamefont {Tsunekawa}\ \emph {et~al.}(2005)\citenamefont
  {Tsunekawa}, \citenamefont {Djayaprawira}, \citenamefont {Nagai},
  \citenamefont {Maehara}, \citenamefont {Yamagata}, \citenamefont {Watanabe},
  \citenamefont {Yuasa}, \citenamefont {Suzuki},\ and\ \citenamefont
  {Ando}}]{4818515}%
  \BibitemOpen
  \bibfield  {author} {\bibinfo {author} {\bibfnamefont {K.}~\bibnamefont
  {Tsunekawa}}, \bibinfo {author} {\bibfnamefont {D.~D.}\ \bibnamefont
  {Djayaprawira}}, \bibinfo {author} {\bibfnamefont {M.}~\bibnamefont {Nagai}},
  \bibinfo {author} {\bibfnamefont {H.}~\bibnamefont {Maehara}}, \bibinfo
  {author} {\bibfnamefont {S.}~\bibnamefont {Yamagata}}, \bibinfo {author}
  {\bibfnamefont {N.}~\bibnamefont {Watanabe}}, \bibinfo {author}
  {\bibfnamefont {S.}~\bibnamefont {Yuasa}}, \bibinfo {author} {\bibfnamefont
  {Y.}~\bibnamefont {Suzuki}}, \ and\ \bibinfo {author} {\bibfnamefont
  {K.}~\bibnamefont {Ando}},\ }\href {\doibase 10.1063/1.2012525} {\bibfield
  {journal} {\bibinfo  {journal} {Applied Physics Letters}\ }\textbf {\bibinfo
  {volume} {87}},\ \bibinfo {pages} {072503 } (\bibinfo {year}
  {2005})}\BibitemShut {NoStop}%
\bibitem [{\citenamefont {Butler}(2008)}]{Butler2008100}%
  \BibitemOpen
  \bibfield  {author} {\bibinfo {author} {\bibfnamefont {W.}~\bibnamefont
  {Butler}},\ }\href {\doibase 10.1088/1468-6996/9/1/014106} {\bibfield
  {journal} {\bibinfo  {journal} {Science and Technology of Advanced
  Materials}\ }\textbf {\bibinfo {volume} {9}},\ \bibinfo {pages} {225 }
  (\bibinfo {year} {2008})}\BibitemShut {NoStop}%
\bibitem [{\citenamefont {Moodera}\ \emph {et~al.}(1995)\citenamefont
  {Moodera}, \citenamefont {Kinder}, \citenamefont {Wong},\ and\ \citenamefont
  {Meservey}}]{PhysRevLett.74.3273}%
  \BibitemOpen
  \bibfield  {author} {\bibinfo {author} {\bibfnamefont {J.~S.}\ \bibnamefont
  {Moodera}}, \bibinfo {author} {\bibfnamefont {L.~R.}\ \bibnamefont {Kinder}},
  \bibinfo {author} {\bibfnamefont {T.~M.}\ \bibnamefont {Wong}}, \ and\
  \bibinfo {author} {\bibfnamefont {R.}~\bibnamefont {Meservey}},\ }\href
  {\doibase 10.1103/PhysRevLett.74.3273} {\bibfield  {journal} {\bibinfo
  {journal} {Phys. Rev. Lett.}\ }\textbf {\bibinfo {volume} {74}},\ \bibinfo
  {pages} {3273} (\bibinfo {year} {1995})}\BibitemShut {NoStop}%
\bibitem [{\citenamefont {Bonell}\ \emph {et~al.}(2012)\citenamefont {Bonell},
  \citenamefont {Hauet}, \citenamefont {Andrieu}, \citenamefont {Bertran},
  \citenamefont {Le~F\`evre}, \citenamefont {Calmels}, \citenamefont {Tejeda},
  \citenamefont {Montaigne}, \citenamefont {Warot-Fonrose}, \citenamefont
  {Belhadji}, \citenamefont {Nicolaou},\ and\ \citenamefont
  {Taleb-Ibrahimi}}]{PhysRevLett.108.176602}%
  \BibitemOpen
  \bibfield  {author} {\bibinfo {author} {\bibfnamefont {F.}~\bibnamefont
  {Bonell}}, \bibinfo {author} {\bibfnamefont {T.}~\bibnamefont {Hauet}},
  \bibinfo {author} {\bibfnamefont {S.}~\bibnamefont {Andrieu}}, \bibinfo
  {author} {\bibfnamefont {F.}~\bibnamefont {Bertran}}, \bibinfo {author}
  {\bibfnamefont {P.}~\bibnamefont {Le~F\`evre}}, \bibinfo {author}
  {\bibfnamefont {L.}~\bibnamefont {Calmels}}, \bibinfo {author} {\bibfnamefont
  {A.}~\bibnamefont {Tejeda}}, \bibinfo {author} {\bibfnamefont
  {F.}~\bibnamefont {Montaigne}}, \bibinfo {author} {\bibfnamefont
  {B.}~\bibnamefont {Warot-Fonrose}}, \bibinfo {author} {\bibfnamefont
  {B.}~\bibnamefont {Belhadji}}, \bibinfo {author} {\bibfnamefont
  {A.}~\bibnamefont {Nicolaou}}, \ and\ \bibinfo {author} {\bibfnamefont
  {A.}~\bibnamefont {Taleb-Ibrahimi}},\ }\href {\doibase
  10.1103/PhysRevLett.108.176602} {\bibfield  {journal} {\bibinfo  {journal}
  {Phys. Rev. Lett.}\ }\textbf {\bibinfo {volume} {108}},\ \bibinfo {pages}
  {176602} (\bibinfo {year} {2012})}\BibitemShut {NoStop}%
\bibitem [{\citenamefont {Lee}\ \emph {et~al.}(2007)\citenamefont {Lee},
  \citenamefont {Hayakawa}, \citenamefont {Ikeda}, \citenamefont {Matsukura},\
  and\ \citenamefont {Ohno}}]{lee:212507}%
  \BibitemOpen
  \bibfield  {author} {\bibinfo {author} {\bibfnamefont {Y.~M.}\ \bibnamefont
  {Lee}}, \bibinfo {author} {\bibfnamefont {J.}~\bibnamefont {Hayakawa}},
  \bibinfo {author} {\bibfnamefont {S.}~\bibnamefont {Ikeda}}, \bibinfo
  {author} {\bibfnamefont {F.}~\bibnamefont {Matsukura}}, \ and\ \bibinfo
  {author} {\bibfnamefont {H.}~\bibnamefont {Ohno}},\ }\href {\doibase
  10.1063/1.2742576} {\bibfield  {journal} {\bibinfo  {journal} {Applied
  Physics Letters}\ }\textbf {\bibinfo {volume} {90}},\ \bibinfo {eid} {212507}
  (\bibinfo {year} {2007})}\BibitemShut {NoStop}%
\bibitem [{\citenamefont {Teixeira}\ \emph {et~al.}(2010)\citenamefont
  {Teixeira}, \citenamefont {Ventura}, \citenamefont {Araujo}, \citenamefont
  {Sousa}, \citenamefont {Wisniowski},\ and\ \citenamefont
  {Freitas}}]{teixeira:262506}%
  \BibitemOpen
  \bibfield  {author} {\bibinfo {author} {\bibfnamefont {J.~M.}\ \bibnamefont
  {Teixeira}}, \bibinfo {author} {\bibfnamefont {J.}~\bibnamefont {Ventura}},
  \bibinfo {author} {\bibfnamefont {J.~P.}\ \bibnamefont {Araujo}}, \bibinfo
  {author} {\bibfnamefont {J.~B.}\ \bibnamefont {Sousa}}, \bibinfo {author}
  {\bibfnamefont {P.}~\bibnamefont {Wisniowski}}, \ and\ \bibinfo {author}
  {\bibfnamefont {P.~P.}\ \bibnamefont {Freitas}},\ }\href {\doibase
  10.1063/1.3458701} {\bibfield  {journal} {\bibinfo  {journal} {Applied
  Physics Letters}\ }\textbf {\bibinfo {volume} {96}},\ \bibinfo {eid} {262506}
  (\bibinfo {year} {2010})}\BibitemShut {NoStop}%
\bibitem [{\citenamefont {Butler}\ \emph {et~al.}(2001)\citenamefont {Butler},
  \citenamefont {Zhang}, \citenamefont {Schulthess},\ and\ \citenamefont
  {MacLaren}}]{PhysRevB.63.054416}%
  \BibitemOpen
  \bibfield  {author} {\bibinfo {author} {\bibfnamefont {W.~H.}\ \bibnamefont
  {Butler}}, \bibinfo {author} {\bibfnamefont {X.-G.}\ \bibnamefont {Zhang}},
  \bibinfo {author} {\bibfnamefont {T.~C.}\ \bibnamefont {Schulthess}}, \ and\
  \bibinfo {author} {\bibfnamefont {J.~M.}\ \bibnamefont {MacLaren}},\ }\href
  {\doibase 10.1103/PhysRevB.63.054416} {\bibfield  {journal} {\bibinfo
  {journal} {Phys. Rev. B}\ }\textbf {\bibinfo {volume} {63}},\ \bibinfo
  {pages} {054416} (\bibinfo {year} {2001})}\BibitemShut {NoStop}%
\bibitem [{\citenamefont {Zhang}\ and\ \citenamefont
  {Butler}(2004)}]{PhysRevB.70.172407}%
  \BibitemOpen
  \bibfield  {author} {\bibinfo {author} {\bibfnamefont {X.-G.}\ \bibnamefont
  {Zhang}}\ and\ \bibinfo {author} {\bibfnamefont {W.~H.}\ \bibnamefont
  {Butler}},\ }\href {\doibase 10.1103/PhysRevB.70.172407} {\bibfield
  {journal} {\bibinfo  {journal} {Phys. Rev. B}\ }\textbf {\bibinfo {volume}
  {70}},\ \bibinfo {pages} {172407} (\bibinfo {year} {2004})}\BibitemShut
  {NoStop}%
\bibitem [{\citenamefont {Wang}\ \emph {et~al.}(2008)\citenamefont {Wang},
  \citenamefont {Han},\ and\ \citenamefont {Zhang}}]{wang:172501}%
  \BibitemOpen
  \bibfield  {author} {\bibinfo {author} {\bibfnamefont {Y.}~\bibnamefont
  {Wang}}, \bibinfo {author} {\bibfnamefont {X.~F.}\ \bibnamefont {Han}}, \
  and\ \bibinfo {author} {\bibfnamefont {X.-G.}\ \bibnamefont {Zhang}},\ }\href
  {\doibase 10.1063/1.3005561} {\bibfield  {journal} {\bibinfo  {journal}
  {Applied Physics Letters}\ }\textbf {\bibinfo {volume} {93}},\ \bibinfo {eid}
  {172501} (\bibinfo {year} {2008})}\BibitemShut {NoStop}%
\bibitem [{\citenamefont {{Yuasa}}\ \emph {et~al.}(2006)\citenamefont
  {{Yuasa}}, \citenamefont {{Fukushima}}, \citenamefont {{Kubota}},
  \citenamefont {{Suzuki}},\ and\ \citenamefont
  {{Ando}}}]{2006ApPhL..89d2505Y}%
  \BibitemOpen
  \bibfield  {author} {\bibinfo {author} {\bibfnamefont {S.}~\bibnamefont
  {{Yuasa}}}, \bibinfo {author} {\bibfnamefont {A.}~\bibnamefont
  {{Fukushima}}}, \bibinfo {author} {\bibfnamefont {H.}~\bibnamefont
  {{Kubota}}}, \bibinfo {author} {\bibfnamefont {Y.}~\bibnamefont {{Suzuki}}},
  \ and\ \bibinfo {author} {\bibfnamefont {K.}~\bibnamefont {{Ando}}},\ }\href
  {\doibase 10.1063/1.2236268} {\bibfield  {journal} {\bibinfo  {journal}
  {Applied Physics Letters}\ }\textbf {\bibinfo {volume} {89}},\ \bibinfo {eid}
  {042505} (\bibinfo {year} {2006})}\BibitemShut {NoStop}%
\bibitem [{\citenamefont {Ikeda}\ \emph {et~al.}(2008)\citenamefont {Ikeda},
  \citenamefont {Hayakawa}, \citenamefont {Ashizawa}, \citenamefont {Lee},
  \citenamefont {Miura}, \citenamefont {Hasegawa}, \citenamefont {Tsunoda},
  \citenamefont {Matsukura},\ and\ \citenamefont {Ohno}}]{4836745}%
  \BibitemOpen
  \bibfield  {author} {\bibinfo {author} {\bibfnamefont {S.}~\bibnamefont
  {Ikeda}}, \bibinfo {author} {\bibfnamefont {J.}~\bibnamefont {Hayakawa}},
  \bibinfo {author} {\bibfnamefont {Y.}~\bibnamefont {Ashizawa}}, \bibinfo
  {author} {\bibfnamefont {Y.~M.}\ \bibnamefont {Lee}}, \bibinfo {author}
  {\bibfnamefont {K.}~\bibnamefont {Miura}}, \bibinfo {author} {\bibfnamefont
  {H.}~\bibnamefont {Hasegawa}}, \bibinfo {author} {\bibfnamefont
  {M.}~\bibnamefont {Tsunoda}}, \bibinfo {author} {\bibfnamefont
  {F.}~\bibnamefont {Matsukura}}, \ and\ \bibinfo {author} {\bibfnamefont
  {H.}~\bibnamefont {Ohno}},\ }\href {\doibase 10.1063/1.2976435} {\bibfield
  {journal} {\bibinfo  {journal} {Applied Physics Letters}\ }\textbf {\bibinfo
  {volume} {93}},\ \bibinfo {pages} {082508 } (\bibinfo {year}
  {2008})}\BibitemShut {NoStop}%
\bibitem [{\citenamefont {Yuasa}\ \emph {et~al.}(2005)\citenamefont {Yuasa},
  \citenamefont {Katayama}, \citenamefont {Nagahama}, \citenamefont
  {Fukushima}, \citenamefont {Kubota}, \citenamefont {Suzuki},\ and\
  \citenamefont {Ando}}]{yuasa:222508}%
  \BibitemOpen
  \bibfield  {author} {\bibinfo {author} {\bibfnamefont {S.}~\bibnamefont
  {Yuasa}}, \bibinfo {author} {\bibfnamefont {T.}~\bibnamefont {Katayama}},
  \bibinfo {author} {\bibfnamefont {T.}~\bibnamefont {Nagahama}}, \bibinfo
  {author} {\bibfnamefont {A.}~\bibnamefont {Fukushima}}, \bibinfo {author}
  {\bibfnamefont {H.}~\bibnamefont {Kubota}}, \bibinfo {author} {\bibfnamefont
  {Y.}~\bibnamefont {Suzuki}}, \ and\ \bibinfo {author} {\bibfnamefont
  {K.}~\bibnamefont {Ando}},\ }\href {\doibase 10.1063/1.2138355} {\bibfield
  {journal} {\bibinfo  {journal} {Applied Physics Letters}\ }\textbf {\bibinfo
  {volume} {87}},\ \bibinfo {eid} {222508} (\bibinfo {year}
  {2005})}\BibitemShut {NoStop}%
\bibitem [{\citenamefont {Zhang}\ and\ \citenamefont
  {Butler}(2003)}]{0953-8984-15-41-R01}%
  \BibitemOpen
  \bibfield  {author} {\bibinfo {author} {\bibfnamefont {X.-G.}\ \bibnamefont
  {Zhang}}\ and\ \bibinfo {author} {\bibfnamefont {W.~H.}\ \bibnamefont
  {Butler}},\ }\href {http://stacks.iop.org/0953-8984/15/i=41/a=R01} {\bibfield
   {journal} {\bibinfo  {journal} {Journal of Physics: Condensed Matter}\
  }\textbf {\bibinfo {volume} {15}},\ \bibinfo {pages} {R1603} (\bibinfo {year}
  {2003})}\BibitemShut {NoStop}%
\bibitem [{\citenamefont {Bischoff}\ \emph {et~al.}(2003)\citenamefont
  {Bischoff}, \citenamefont {Yamada}, \citenamefont {Fang}, \citenamefont
  {de~Groot},\ and\ \citenamefont {van Kempen}}]{PhysRevB.68.045422}%
  \BibitemOpen
  \bibfield  {author} {\bibinfo {author} {\bibfnamefont {M.~M.~J.}\
  \bibnamefont {Bischoff}}, \bibinfo {author} {\bibfnamefont {T.~K.}\
  \bibnamefont {Yamada}}, \bibinfo {author} {\bibfnamefont {C.~M.}\
  \bibnamefont {Fang}}, \bibinfo {author} {\bibfnamefont {R.~A.}\ \bibnamefont
  {de~Groot}}, \ and\ \bibinfo {author} {\bibfnamefont {H.}~\bibnamefont {van
  Kempen}},\ }\href {\doibase 10.1103/PhysRevB.68.045422} {\bibfield  {journal}
  {\bibinfo  {journal} {Phys. Rev. B}\ }\textbf {\bibinfo {volume} {68}},\
  \bibinfo {pages} {045422} (\bibinfo {year} {2003})}\BibitemShut {NoStop}%
\bibitem [{\citenamefont {Plucinski}\ \emph {et~al.}(2009)\citenamefont
  {Plucinski}, \citenamefont {Zhao}, \citenamefont {Schneider}, \citenamefont
  {Sinkovic},\ and\ \citenamefont {Vescovo}}]{PhysRevB.80.184430}%
  \BibitemOpen
  \bibfield  {author} {\bibinfo {author} {\bibfnamefont {L.}~\bibnamefont
  {Plucinski}}, \bibinfo {author} {\bibfnamefont {Y.}~\bibnamefont {Zhao}},
  \bibinfo {author} {\bibfnamefont {C.~M.}\ \bibnamefont {Schneider}}, \bibinfo
  {author} {\bibfnamefont {B.}~\bibnamefont {Sinkovic}}, \ and\ \bibinfo
  {author} {\bibfnamefont {E.}~\bibnamefont {Vescovo}},\ }\href {\doibase
  10.1103/PhysRevB.80.184430} {\bibfield  {journal} {\bibinfo  {journal} {Phys.
  Rev. B}\ }\textbf {\bibinfo {volume} {80}},\ \bibinfo {pages} {184430}
  (\bibinfo {year} {2009})}\BibitemShut {NoStop}%
\bibitem [{\citenamefont {Tiusan}\ \emph {et~al.}(2004)\citenamefont {Tiusan},
  \citenamefont {Faure-Vincent}, \citenamefont {Bellouard}, \citenamefont
  {Hehn}, \citenamefont {Jouguelet},\ and\ \citenamefont
  {Schuhl}}]{ISI:000223717600046}%
  \BibitemOpen
  \bibfield  {author} {\bibinfo {author} {\bibfnamefont {C.}~\bibnamefont
  {Tiusan}}, \bibinfo {author} {\bibfnamefont {J.}~\bibnamefont
  {Faure-Vincent}}, \bibinfo {author} {\bibfnamefont {C.}~\bibnamefont
  {Bellouard}}, \bibinfo {author} {\bibfnamefont {M.}~\bibnamefont {Hehn}},
  \bibinfo {author} {\bibfnamefont {E.}~\bibnamefont {Jouguelet}}, \ and\
  \bibinfo {author} {\bibfnamefont {A.}~\bibnamefont {Schuhl}},\ }\href
  {\doibase 10.1103/PhysRevLett.93.106602} {\bibfield  {journal} {\bibinfo
  {journal} {Physical Review Letters}\ }\textbf {\bibinfo {volume} {93}},\
  \bibinfo {pages} {106602} (\bibinfo {year} {2004})}\BibitemShut {NoStop}%
\bibitem [{\citenamefont {Nordheim}(1931)}]{nordheim1931elektronentheorie}%
  \BibitemOpen
  \bibfield  {author} {\bibinfo {author} {\bibfnamefont {L.}~\bibnamefont
  {Nordheim}},\ }\href@noop {} {\bibfield  {journal} {\bibinfo  {journal}
  {Annalen der Physik}\ }\textbf {\bibinfo {volume} {401}},\ \bibinfo {pages}
  {607} (\bibinfo {year} {1931})}\BibitemShut {NoStop}%
\bibitem [{\citenamefont {Bellaiche}\ and\ \citenamefont
  {Vanderbilt}(2000)}]{PhysRevB.61.7877}%
  \BibitemOpen
  \bibfield  {author} {\bibinfo {author} {\bibfnamefont {L.}~\bibnamefont
  {Bellaiche}}\ and\ \bibinfo {author} {\bibfnamefont {D.}~\bibnamefont
  {Vanderbilt}},\ }\href {\doibase 10.1103/PhysRevB.61.7877} {\bibfield
  {journal} {\bibinfo  {journal} {Phys. Rev. B}\ }\textbf {\bibinfo {volume}
  {61}},\ \bibinfo {pages} {7877} (\bibinfo {year} {2000})}\BibitemShut
  {NoStop}%
\bibitem [{\citenamefont {Soven}(1967)}]{soven1967coherent}%
  \BibitemOpen
  \bibfield  {author} {\bibinfo {author} {\bibfnamefont {P.}~\bibnamefont
  {Soven}},\ }\href@noop {} {\bibfield  {journal} {\bibinfo  {journal}
  {Physical Review}\ }\textbf {\bibinfo {volume} {156}},\ \bibinfo {pages}
  {809} (\bibinfo {year} {1967})}\BibitemShut {NoStop}%
\bibitem [{\citenamefont {MacLaren}\ \emph {et~al.}(1989)\citenamefont
  {MacLaren}, \citenamefont {Crampin}, \citenamefont {Vvedensky},\ and\
  \citenamefont {Pendry}}]{PhysRevB.40.12164}%
  \BibitemOpen
  \bibfield  {author} {\bibinfo {author} {\bibfnamefont {J.~M.}\ \bibnamefont
  {MacLaren}}, \bibinfo {author} {\bibfnamefont {S.}~\bibnamefont {Crampin}},
  \bibinfo {author} {\bibfnamefont {D.~D.}\ \bibnamefont {Vvedensky}}, \ and\
  \bibinfo {author} {\bibfnamefont {J.~B.}\ \bibnamefont {Pendry}},\ }\href
  {\doibase 10.1103/PhysRevB.40.12164} {\bibfield  {journal} {\bibinfo
  {journal} {Phys. Rev. B}\ }\textbf {\bibinfo {volume} {40}},\ \bibinfo
  {pages} {12164} (\bibinfo {year} {1989})}\BibitemShut {NoStop}%
\bibitem [{\citenamefont {Butler}(1985)}]{PhysRevB.31.3260}%
  \BibitemOpen
  \bibfield  {author} {\bibinfo {author} {\bibfnamefont {W.~H.}\ \bibnamefont
  {Butler}},\ }\href {\doibase 10.1103/PhysRevB.31.3260} {\bibfield  {journal}
  {\bibinfo  {journal} {Phys. Rev. B}\ }\textbf {\bibinfo {volume} {31}},\
  \bibinfo {pages} {3260} (\bibinfo {year} {1985})}\BibitemShut {NoStop}%
\bibitem [{\citenamefont {Faulkner}(2010)}]{faulkner201050}%
  \BibitemOpen
  \bibfield  {author} {\bibinfo {author} {\bibfnamefont {J.}~\bibnamefont
  {Faulkner}},\ }\href@noop {} {\bibfield  {journal} {\bibinfo  {journal}
  {Molecular Physics}\ }\textbf {\bibinfo {volume} {108}},\ \bibinfo {pages}
  {3189} (\bibinfo {year} {2010})}\BibitemShut {NoStop}%
\bibitem [{\citenamefont {Marzari}\ \emph {et~al.}(1994)\citenamefont
  {Marzari}, \citenamefont {de~Gironcoli},\ and\ \citenamefont
  {Baroni}}]{PhysRevLett.72.4001}%
  \BibitemOpen
  \bibfield  {author} {\bibinfo {author} {\bibfnamefont {N.}~\bibnamefont
  {Marzari}}, \bibinfo {author} {\bibfnamefont {S.}~\bibnamefont
  {de~Gironcoli}}, \ and\ \bibinfo {author} {\bibfnamefont {S.}~\bibnamefont
  {Baroni}},\ }\href {\doibase 10.1103/PhysRevLett.72.4001} {\bibfield
  {journal} {\bibinfo  {journal} {Phys. Rev. Lett.}\ }\textbf {\bibinfo
  {volume} {72}},\ \bibinfo {pages} {4001} (\bibinfo {year}
  {1994})}\BibitemShut {NoStop}%
\bibitem [{\citenamefont {Giannozzi}\ \emph {et~al.}(2009)\citenamefont
  {Giannozzi}, \citenamefont {Baroni}, \citenamefont {Bonini}, \citenamefont
  {Calandra}, \citenamefont {Car}, \citenamefont {Cavazzoni}, \citenamefont
  {Ceresoli}, \citenamefont {Chiarotti}, \citenamefont {Cococcioni},
  \citenamefont {Dabo}, \citenamefont {Corso}, \citenamefont {de~Gironcoli},
  \citenamefont {Fabris}, \citenamefont {Fratesi}, \citenamefont {Gebauer},
  \citenamefont {Gerstmann}, \citenamefont {Gougoussis}, \citenamefont
  {Kokalj}, \citenamefont {Lazzeri}, \citenamefont {Martin-Samos},
  \citenamefont {Marzari}, \citenamefont {Mauri}, \citenamefont {Mazzarello},
  \citenamefont {Paolini}, \citenamefont {Pasquarello}, \citenamefont
  {Paulatto}, \citenamefont {Sbraccia}, \citenamefont {Scandolo}, \citenamefont
  {Sclauzero}, \citenamefont {Seitsonen}, \citenamefont {Smogunov},
  \citenamefont {Umari},\ and\ \citenamefont
  {Wentzcovitch}}]{0953-8984-21-39-395502}%
  \BibitemOpen
  \bibfield  {author} {\bibinfo {author} {\bibfnamefont {P.}~\bibnamefont
  {Giannozzi}}, \bibinfo {author} {\bibfnamefont {S.}~\bibnamefont {Baroni}},
  \bibinfo {author} {\bibfnamefont {N.}~\bibnamefont {Bonini}}, \bibinfo
  {author} {\bibfnamefont {M.}~\bibnamefont {Calandra}}, \bibinfo {author}
  {\bibfnamefont {R.}~\bibnamefont {Car}}, \bibinfo {author} {\bibfnamefont
  {C.}~\bibnamefont {Cavazzoni}}, \bibinfo {author} {\bibfnamefont
  {D.}~\bibnamefont {Ceresoli}}, \bibinfo {author} {\bibfnamefont {G.~L.}\
  \bibnamefont {Chiarotti}}, \bibinfo {author} {\bibfnamefont {M.}~\bibnamefont
  {Cococcioni}}, \bibinfo {author} {\bibfnamefont {I.}~\bibnamefont {Dabo}},
  \bibinfo {author} {\bibfnamefont {A.~D.}\ \bibnamefont {Corso}}, \bibinfo
  {author} {\bibfnamefont {S.}~\bibnamefont {de~Gironcoli}}, \bibinfo {author}
  {\bibfnamefont {S.}~\bibnamefont {Fabris}}, \bibinfo {author} {\bibfnamefont
  {G.}~\bibnamefont {Fratesi}}, \bibinfo {author} {\bibfnamefont
  {R.}~\bibnamefont {Gebauer}}, \bibinfo {author} {\bibfnamefont
  {U.}~\bibnamefont {Gerstmann}}, \bibinfo {author} {\bibfnamefont
  {C.}~\bibnamefont {Gougoussis}}, \bibinfo {author} {\bibfnamefont
  {A.}~\bibnamefont {Kokalj}}, \bibinfo {author} {\bibfnamefont
  {M.}~\bibnamefont {Lazzeri}}, \bibinfo {author} {\bibfnamefont
  {L.}~\bibnamefont {Martin-Samos}}, \bibinfo {author} {\bibfnamefont
  {N.}~\bibnamefont {Marzari}}, \bibinfo {author} {\bibfnamefont
  {F.}~\bibnamefont {Mauri}}, \bibinfo {author} {\bibfnamefont
  {R.}~\bibnamefont {Mazzarello}}, \bibinfo {author} {\bibfnamefont
  {S.}~\bibnamefont {Paolini}}, \bibinfo {author} {\bibfnamefont
  {A.}~\bibnamefont {Pasquarello}}, \bibinfo {author} {\bibfnamefont
  {L.}~\bibnamefont {Paulatto}}, \bibinfo {author} {\bibfnamefont
  {C.}~\bibnamefont {Sbraccia}}, \bibinfo {author} {\bibfnamefont
  {S.}~\bibnamefont {Scandolo}}, \bibinfo {author} {\bibfnamefont
  {G.}~\bibnamefont {Sclauzero}}, \bibinfo {author} {\bibfnamefont {A.~P.}\
  \bibnamefont {Seitsonen}}, \bibinfo {author} {\bibfnamefont {A.}~\bibnamefont
  {Smogunov}}, \bibinfo {author} {\bibfnamefont {P.}~\bibnamefont {Umari}}, \
  and\ \bibinfo {author} {\bibfnamefont {R.~M.}\ \bibnamefont {Wentzcovitch}},\
  }\href {http://stacks.iop.org/0953-8984/21/i=39/a=395502} {\bibfield
  {journal} {\bibinfo  {journal} {Journal of Physics: Condensed Matter}\
  }\textbf {\bibinfo {volume} {21}},\ \bibinfo {pages} {395502} (\bibinfo
  {year} {2009})}\BibitemShut {NoStop}%
\bibitem [{\citenamefont {Monkhorst}\ and\ \citenamefont
  {Pack}(1976)}]{PhysRevB.13.5188}%
  \BibitemOpen
  \bibfield  {author} {\bibinfo {author} {\bibfnamefont {H.~J.}\ \bibnamefont
  {Monkhorst}}\ and\ \bibinfo {author} {\bibfnamefont {J.~D.}\ \bibnamefont
  {Pack}},\ }\href {\doibase 10.1103/PhysRevB.13.5188} {\bibfield  {journal}
  {\bibinfo  {journal} {Phys. Rev. B}\ }\textbf {\bibinfo {volume} {13}},\
  \bibinfo {pages} {5188} (\bibinfo {year} {1976})}\BibitemShut {NoStop}%
\bibitem [{\citenamefont {Bose}\ \emph {et~al.}(2010)\citenamefont {Bose},
  \citenamefont {Zahn}, \citenamefont {Henk},\ and\ \citenamefont
  {Mertig}}]{bose2010tailoring}%
  \BibitemOpen
  \bibfield  {author} {\bibinfo {author} {\bibfnamefont {P.}~\bibnamefont
  {Bose}}, \bibinfo {author} {\bibfnamefont {P.}~\bibnamefont {Zahn}}, \bibinfo
  {author} {\bibfnamefont {J.}~\bibnamefont {Henk}}, \ and\ \bibinfo {author}
  {\bibfnamefont {I.}~\bibnamefont {Mertig}},\ }\href@noop {} {\bibfield
  {journal} {\bibinfo  {journal} {Physical Review B}\ }\textbf {\bibinfo
  {volume} {82}},\ \bibinfo {pages} {014412} (\bibinfo {year}
  {2010})}\BibitemShut {NoStop}%
\bibitem [{\citenamefont {Miura}\ \emph {et~al.}(2012)\citenamefont {Miura},
  \citenamefont {Muramoto}, \citenamefont {Abe},\ and\ \citenamefont
  {Shirai}}]{miura2012first}%
  \BibitemOpen
  \bibfield  {author} {\bibinfo {author} {\bibfnamefont {Y.}~\bibnamefont
  {Miura}}, \bibinfo {author} {\bibfnamefont {S.}~\bibnamefont {Muramoto}},
  \bibinfo {author} {\bibfnamefont {K.}~\bibnamefont {Abe}}, \ and\ \bibinfo
  {author} {\bibfnamefont {M.}~\bibnamefont {Shirai}},\ }\href@noop {}
  {\bibfield  {journal} {\bibinfo  {journal} {Physical Review B}\ }\textbf
  {\bibinfo {volume} {86}},\ \bibinfo {pages} {024426} (\bibinfo {year}
  {2012})}\BibitemShut {NoStop}%
\bibitem [{\citenamefont {Wang}\ \emph
  {et~al.}(2010{\natexlab{a}})\citenamefont {Wang}, \citenamefont {Li},
  \citenamefont {Liu}, \citenamefont {Liu},\ and\ \citenamefont
  {Xing}}]{wang:123906}%
  \BibitemOpen
  \bibfield  {author} {\bibinfo {author} {\bibfnamefont {W.}~\bibnamefont
  {Wang}}, \bibinfo {author} {\bibfnamefont {B.}~\bibnamefont {Li}}, \bibinfo
  {author} {\bibfnamefont {S.}~\bibnamefont {Liu}}, \bibinfo {author}
  {\bibfnamefont {M.}~\bibnamefont {Liu}}, \ and\ \bibinfo {author}
  {\bibfnamefont {Z.~W.}\ \bibnamefont {Xing}},\ }\href {\doibase
  10.1063/1.3448233} {\bibfield  {journal} {\bibinfo  {journal} {Journal of
  Applied Physics}\ }\textbf {\bibinfo {volume} {107}},\ \bibinfo {eid}
  {123906} (\bibinfo {year} {2010}{\natexlab{a}})}\BibitemShut {NoStop}%
\bibitem [{\citenamefont {Victora}\ and\ \citenamefont
  {Falicov}(1984)}]{PhysRevB.30.259}%
  \BibitemOpen
  \bibfield  {author} {\bibinfo {author} {\bibfnamefont {R.~H.}\ \bibnamefont
  {Victora}}\ and\ \bibinfo {author} {\bibfnamefont {L.~M.}\ \bibnamefont
  {Falicov}},\ }\href {\doibase 10.1103/PhysRevB.30.259} {\bibfield  {journal}
  {\bibinfo  {journal} {Phys. Rev. B}\ }\textbf {\bibinfo {volume} {30}},\
  \bibinfo {pages} {259} (\bibinfo {year} {1984})}\BibitemShut {NoStop}%
\bibitem [{\citenamefont {Soderlind}\ \emph {et~al.}(1992)\citenamefont
  {Soderlind}, \citenamefont {Johansson},\ and\ \citenamefont
  {Eriksson}}]{Söderlind19922037}%
  \BibitemOpen
  \bibfield  {author} {\bibinfo {author} {\bibfnamefont {P.}~\bibnamefont
  {Soderlind}}, \bibinfo {author} {\bibfnamefont {B.}~\bibnamefont
  {Johansson}}, \ and\ \bibinfo {author} {\bibfnamefont {O.}~\bibnamefont
  {Eriksson}},\ }\href {\doibase 10.1016/0304-8853(92)91658-G} {\bibfield
  {journal} {\bibinfo  {journal} {Journal of Magnetism and Magnetic Materials}\
  }\textbf {\bibinfo {volume} {104–107, Part 3}},\ \bibinfo {pages} {2037 }
  (\bibinfo {year} {1992})}\BibitemShut {NoStop}%
\bibitem [{\citenamefont {{Bozorth}}(1993)}]{1993ferr.book.....B}%
  \BibitemOpen
  \bibfield  {author} {\bibinfo {author} {\bibfnamefont {R.~M.}\ \bibnamefont
  {{Bozorth}}},\ }\href@noop {} {\emph {\bibinfo {title} {Ferromagnetism , by
  Richard M.~Bozorth, pp.~992.~ISBN 0-7803-1032-2.~Wiley-VCH , August 1993.}}}\
  (\bibinfo {year} {1993})\BibitemShut {NoStop}%
\bibitem [{\citenamefont {MacLaren}\ \emph {et~al.}(1999)\citenamefont
  {MacLaren}, \citenamefont {Zhang}, \citenamefont {Butler},\ and\
  \citenamefont {Wang}}]{PhysRevB.59.5470}%
  \BibitemOpen
  \bibfield  {author} {\bibinfo {author} {\bibfnamefont {J.~M.}\ \bibnamefont
  {MacLaren}}, \bibinfo {author} {\bibfnamefont {X.-G.}\ \bibnamefont {Zhang}},
  \bibinfo {author} {\bibfnamefont {W.~H.}\ \bibnamefont {Butler}}, \ and\
  \bibinfo {author} {\bibfnamefont {X.}~\bibnamefont {Wang}},\ }\href {\doibase
  10.1103/PhysRevB.59.5470} {\bibfield  {journal} {\bibinfo  {journal} {Phys.
  Rev. B}\ }\textbf {\bibinfo {volume} {59}},\ \bibinfo {pages} {5470}
  (\bibinfo {year} {1999})}\BibitemShut {NoStop}%
\bibitem [{\citenamefont {Richter}\ and\ \citenamefont
  {Eschrig}(1988)}]{0305-4608-18-8-017}%
  \BibitemOpen
  \bibfield  {author} {\bibinfo {author} {\bibfnamefont {R.}~\bibnamefont
  {Richter}}\ and\ \bibinfo {author} {\bibfnamefont {H.}~\bibnamefont
  {Eschrig}},\ }\href {http://stacks.iop.org/0305-4608/18/i=8/a=017} {\bibfield
   {journal} {\bibinfo  {journal} {Journal of Physics F: Metal Physics}\
  }\textbf {\bibinfo {volume} {18}},\ \bibinfo {pages} {1813} (\bibinfo {year}
  {1988})}\BibitemShut {NoStop}%
\bibitem [{\citenamefont {Moriyama}\ \emph {et~al.}(2006)\citenamefont
  {Moriyama}, \citenamefont {Ni}, \citenamefont {Wang}, \citenamefont {Zhang},\
  and\ \citenamefont {Xiao}}]{moriyama:222503}%
  \BibitemOpen
  \bibfield  {author} {\bibinfo {author} {\bibfnamefont {T.}~\bibnamefont
  {Moriyama}}, \bibinfo {author} {\bibfnamefont {C.}~\bibnamefont {Ni}},
  \bibinfo {author} {\bibfnamefont {W.~G.}\ \bibnamefont {Wang}}, \bibinfo
  {author} {\bibfnamefont {X.}~\bibnamefont {Zhang}}, \ and\ \bibinfo {author}
  {\bibfnamefont {J.~Q.}\ \bibnamefont {Xiao}},\ }\href {\doibase
  10.1063/1.2207835} {\bibfield  {journal} {\bibinfo  {journal} {Applied
  Physics Letters}\ }\textbf {\bibinfo {volume} {88}},\ \bibinfo {eid} {222503}
  (\bibinfo {year} {2006})}\BibitemShut {NoStop}%
\bibitem [{\citenamefont {Shikada}\ \emph {et~al.}(2009)\citenamefont
  {Shikada}, \citenamefont {Ohtake}, \citenamefont {Kirino},\ and\
  \citenamefont {Futamoto}}]{shikada:07C303}%
  \BibitemOpen
  \bibfield  {author} {\bibinfo {author} {\bibfnamefont {K.}~\bibnamefont
  {Shikada}}, \bibinfo {author} {\bibfnamefont {M.}~\bibnamefont {Ohtake}},
  \bibinfo {author} {\bibfnamefont {F.}~\bibnamefont {Kirino}}, \ and\ \bibinfo
  {author} {\bibfnamefont {M.}~\bibnamefont {Futamoto}},\ }\href {\doibase
  10.1063/1.3067854} {\bibfield  {journal} {\bibinfo  {journal} {Journal of
  Applied Physics}\ }\textbf {\bibinfo {volume} {105}},\ \bibinfo {eid}
  {07C303} (\bibinfo {year} {2009})}\BibitemShut {NoStop}%
\bibitem [{\citenamefont {Burton}\ \emph {et~al.}(2006)\citenamefont {Burton},
  \citenamefont {Jaswal}, \citenamefont {Tsymbal}, \citenamefont {Mryasov},\
  and\ \citenamefont {Heinonen}}]{burton:142507}%
  \BibitemOpen
  \bibfield  {author} {\bibinfo {author} {\bibfnamefont {J.~D.}\ \bibnamefont
  {Burton}}, \bibinfo {author} {\bibfnamefont {S.~S.}\ \bibnamefont {Jaswal}},
  \bibinfo {author} {\bibfnamefont {E.~Y.}\ \bibnamefont {Tsymbal}}, \bibinfo
  {author} {\bibfnamefont {O.~N.}\ \bibnamefont {Mryasov}}, \ and\ \bibinfo
  {author} {\bibfnamefont {O.~G.}\ \bibnamefont {Heinonen}},\ }\href {\doibase
  10.1063/1.2360189} {\bibfield  {journal} {\bibinfo  {journal} {Applied
  Physics Letters}\ }\textbf {\bibinfo {volume} {89}},\ \bibinfo {eid} {142507}
  (\bibinfo {year} {2006})}\BibitemShut {NoStop}%
\bibitem [{\citenamefont {Smogunov}\ \emph {et~al.}(2004)\citenamefont
  {Smogunov}, \citenamefont {Dal~Corso},\ and\ \citenamefont
  {Tosatti}}]{PhysRevB.70.045417}%
  \BibitemOpen
  \bibfield  {author} {\bibinfo {author} {\bibfnamefont {A.}~\bibnamefont
  {Smogunov}}, \bibinfo {author} {\bibfnamefont {A.}~\bibnamefont {Dal~Corso}},
  \ and\ \bibinfo {author} {\bibfnamefont {E.}~\bibnamefont {Tosatti}},\ }\href
  {\doibase 10.1103/PhysRevB.70.045417} {\bibfield  {journal} {\bibinfo
  {journal} {Phys. Rev. B}\ }\textbf {\bibinfo {volume} {70}},\ \bibinfo
  {pages} {045417} (\bibinfo {year} {2004})}\BibitemShut {NoStop}%
\bibitem [{\citenamefont {Wang}\ \emph
  {et~al.}(2010{\natexlab{b}})\citenamefont {Wang}, \citenamefont {Zhang},
  \citenamefont {Zhang}, \citenamefont {Cheng},\ and\ \citenamefont
  {Han}}]{PhysRevB.82.054405}%
  \BibitemOpen
  \bibfield  {author} {\bibinfo {author} {\bibfnamefont {Y.}~\bibnamefont
  {Wang}}, \bibinfo {author} {\bibfnamefont {J.}~\bibnamefont {Zhang}},
  \bibinfo {author} {\bibfnamefont {X.-G.}\ \bibnamefont {Zhang}}, \bibinfo
  {author} {\bibfnamefont {H.-P.}\ \bibnamefont {Cheng}}, \ and\ \bibinfo
  {author} {\bibfnamefont {X.~F.}\ \bibnamefont {Han}},\ }\href {\doibase
  10.1103/PhysRevB.82.054405} {\bibfield  {journal} {\bibinfo  {journal} {Phys.
  Rev. B}\ }\textbf {\bibinfo {volume} {82}},\ \bibinfo {pages} {054405}
  (\bibinfo {year} {2010}{\natexlab{b}})}\BibitemShut {NoStop}%
\bibitem [{\citenamefont {Stroscio}\ \emph {et~al.}(1995)\citenamefont
  {Stroscio}, \citenamefont {Pierce}, \citenamefont {Davies}, \citenamefont
  {Celotta},\ and\ \citenamefont {Weinert}}]{PhysRevLett.75.2960}%
  \BibitemOpen
  \bibfield  {author} {\bibinfo {author} {\bibfnamefont {J.~A.}\ \bibnamefont
  {Stroscio}}, \bibinfo {author} {\bibfnamefont {D.~T.}\ \bibnamefont
  {Pierce}}, \bibinfo {author} {\bibfnamefont {A.}~\bibnamefont {Davies}},
  \bibinfo {author} {\bibfnamefont {R.~J.}\ \bibnamefont {Celotta}}, \ and\
  \bibinfo {author} {\bibfnamefont {M.}~\bibnamefont {Weinert}},\ }\href
  {\doibase 10.1103/PhysRevLett.75.2960} {\bibfield  {journal} {\bibinfo
  {journal} {Phys. Rev. Lett.}\ }\textbf {\bibinfo {volume} {75}},\ \bibinfo
  {pages} {2960} (\bibinfo {year} {1995})}\BibitemShut {NoStop}%
\bibitem [{\citenamefont {Tiusan}\ \emph {et~al.}(2006)\citenamefont {Tiusan},
  \citenamefont {Faure-Vincent}, \citenamefont {Sicot}, \citenamefont {Hehn},
  \citenamefont {Bellouard}, \citenamefont {Montaigne}, \citenamefont
  {Andrieu},\ and\ \citenamefont {Schuhl}}]{Tiusan2006112}%
  \BibitemOpen
  \bibfield  {author} {\bibinfo {author} {\bibfnamefont {C.}~\bibnamefont
  {Tiusan}}, \bibinfo {author} {\bibfnamefont {J.}~\bibnamefont
  {Faure-Vincent}}, \bibinfo {author} {\bibfnamefont {M.}~\bibnamefont
  {Sicot}}, \bibinfo {author} {\bibfnamefont {M.}~\bibnamefont {Hehn}},
  \bibinfo {author} {\bibfnamefont {C.}~\bibnamefont {Bellouard}}, \bibinfo
  {author} {\bibfnamefont {F.}~\bibnamefont {Montaigne}}, \bibinfo {author}
  {\bibfnamefont {S.}~\bibnamefont {Andrieu}}, \ and\ \bibinfo {author}
  {\bibfnamefont {A.}~\bibnamefont {Schuhl}},\ }\href {\doibase
  10.1016/j.mseb.2005.09.054} {\bibfield  {journal} {\bibinfo  {journal}
  {Materials Science and Engineering: B}\ }\textbf {\bibinfo {volume} {126}},\
  \bibinfo {pages} {112 } (\bibinfo {year} {2006})},\ \bibinfo {note} {eMRS
  2005, Symposium B, Spintronics}\BibitemShut {NoStop}%
\bibitem [{\citenamefont {Yuasa}\ \emph {et~al.}(2004)\citenamefont {Yuasa},
  \citenamefont {Nagahama}, \citenamefont {Fukushima}, \citenamefont {Suzuki},\
  and\ \citenamefont {Ando}}]{yuasa2004giant}%
  \BibitemOpen
  \bibfield  {author} {\bibinfo {author} {\bibfnamefont {S.}~\bibnamefont
  {Yuasa}}, \bibinfo {author} {\bibfnamefont {T.}~\bibnamefont {Nagahama}},
  \bibinfo {author} {\bibfnamefont {A.}~\bibnamefont {Fukushima}}, \bibinfo
  {author} {\bibfnamefont {Y.}~\bibnamefont {Suzuki}}, \ and\ \bibinfo {author}
  {\bibfnamefont {K.}~\bibnamefont {Ando}},\ }\href@noop {} {\bibfield
  {journal} {\bibinfo  {journal} {Nature materials}\ }\textbf {\bibinfo
  {volume} {3}},\ \bibinfo {pages} {868} (\bibinfo {year} {2004})}\BibitemShut
  {NoStop}%
\bibitem [{\citenamefont {Hayakawa}\ \emph {et~al.}(2005)\citenamefont
  {Hayakawa}, \citenamefont {Ikeda}, \citenamefont {Matsukura}, \citenamefont
  {Takahashi},\ and\ \citenamefont {Ohno}}]{hayakawa2005dependence}%
  \BibitemOpen
  \bibfield  {author} {\bibinfo {author} {\bibfnamefont {J.}~\bibnamefont
  {Hayakawa}}, \bibinfo {author} {\bibfnamefont {S.}~\bibnamefont {Ikeda}},
  \bibinfo {author} {\bibfnamefont {F.}~\bibnamefont {Matsukura}}, \bibinfo
  {author} {\bibfnamefont {H.}~\bibnamefont {Takahashi}}, \ and\ \bibinfo
  {author} {\bibfnamefont {H.}~\bibnamefont {Ohno}},\ }\href@noop {} {\bibfield
   {journal} {\bibinfo  {journal} {arXiv preprint cond-mat/0504051}\ }
  (\bibinfo {year} {2005})}\BibitemShut {NoStop}%
\bibitem [{\citenamefont {Wang}\ \emph {et~al.}(2006)\citenamefont {Wang},
  \citenamefont {Lu}, \citenamefont {Zhang},\ and\ \citenamefont
  {Han}}]{PhysRevLett.97.087210}%
  \BibitemOpen
  \bibfield  {author} {\bibinfo {author} {\bibfnamefont {Y.}~\bibnamefont
  {Wang}}, \bibinfo {author} {\bibfnamefont {Z.-Y.}\ \bibnamefont {Lu}},
  \bibinfo {author} {\bibfnamefont {X.-G.}\ \bibnamefont {Zhang}}, \ and\
  \bibinfo {author} {\bibfnamefont {X.~F.}\ \bibnamefont {Han}},\ }\href
  {\doibase 10.1103/PhysRevLett.97.087210} {\bibfield  {journal} {\bibinfo
  {journal} {Phys. Rev. Lett.}\ }\textbf {\bibinfo {volume} {97}},\ \bibinfo
  {pages} {087210} (\bibinfo {year} {2006})}\BibitemShut {NoStop}%
\bibitem [{\citenamefont {Meyerheim}\ \emph {et~al.}(2001)\citenamefont
  {Meyerheim}, \citenamefont {Popescu}, \citenamefont {Kirschner},
  \citenamefont {Jedrecy}, \citenamefont {Sauvage-Simkin}, \citenamefont
  {Heinrich},\ and\ \citenamefont {Pinchaux}}]{PhysRevLett.87.076102}%
  \BibitemOpen
  \bibfield  {author} {\bibinfo {author} {\bibfnamefont {H.~L.}\ \bibnamefont
  {Meyerheim}}, \bibinfo {author} {\bibfnamefont {R.}~\bibnamefont {Popescu}},
  \bibinfo {author} {\bibfnamefont {J.}~\bibnamefont {Kirschner}}, \bibinfo
  {author} {\bibfnamefont {N.}~\bibnamefont {Jedrecy}}, \bibinfo {author}
  {\bibfnamefont {M.}~\bibnamefont {Sauvage-Simkin}}, \bibinfo {author}
  {\bibfnamefont {B.}~\bibnamefont {Heinrich}}, \ and\ \bibinfo {author}
  {\bibfnamefont {R.}~\bibnamefont {Pinchaux}},\ }\href {\doibase
  10.1103/PhysRevLett.87.076102} {\bibfield  {journal} {\bibinfo  {journal}
  {Phys. Rev. Lett.}\ }\textbf {\bibinfo {volume} {87}},\ \bibinfo {pages}
  {076102} (\bibinfo {year} {2001})}\BibitemShut {NoStop}%
\bibitem [{\citenamefont {Zhang}\ \emph {et~al.}(2003)\citenamefont {Zhang},
  \citenamefont {Butler},\ and\ \citenamefont
  {Bandyopadhyay}}]{PhysRevB.68.092402}%
  \BibitemOpen
  \bibfield  {author} {\bibinfo {author} {\bibfnamefont {X.-G.}\ \bibnamefont
  {Zhang}}, \bibinfo {author} {\bibfnamefont {W.~H.}\ \bibnamefont {Butler}}, \
  and\ \bibinfo {author} {\bibfnamefont {A.}~\bibnamefont {Bandyopadhyay}},\
  }\href {\doibase 10.1103/PhysRevB.68.092402} {\bibfield  {journal} {\bibinfo
  {journal} {Phys. Rev. B}\ }\textbf {\bibinfo {volume} {68}},\ \bibinfo
  {pages} {092402} (\bibinfo {year} {2003})}\BibitemShut {NoStop}%
\bibitem [{\citenamefont {Tiusan}\ \emph {et~al.}(2007)\citenamefont {Tiusan},
  \citenamefont {Greullet}, \citenamefont {Hehn}, \citenamefont {Montaigne},
  \citenamefont {Andrieu},\ and\ \citenamefont
  {Schuhl}}]{0953-8984-19-16-165201}%
  \BibitemOpen
  \bibfield  {author} {\bibinfo {author} {\bibfnamefont {C.}~\bibnamefont
  {Tiusan}}, \bibinfo {author} {\bibfnamefont {F.}~\bibnamefont {Greullet}},
  \bibinfo {author} {\bibfnamefont {M.}~\bibnamefont {Hehn}}, \bibinfo {author}
  {\bibfnamefont {F.}~\bibnamefont {Montaigne}}, \bibinfo {author}
  {\bibfnamefont {S.}~\bibnamefont {Andrieu}}, \ and\ \bibinfo {author}
  {\bibfnamefont {A.}~\bibnamefont {Schuhl}},\ }\href
  {http://stacks.iop.org/0953-8984/19/i=16/a=165201} {\bibfield  {journal}
  {\bibinfo  {journal} {Journal of Physics: Condensed Matter}\ }\textbf
  {\bibinfo {volume} {19}},\ \bibinfo {pages} {165201} (\bibinfo {year}
  {2007})}\BibitemShut {NoStop}%
\bibitem [{\citenamefont {Belashchenko}\ \emph {et~al.}(2005)\citenamefont
  {Belashchenko}, \citenamefont {Velev},\ and\ \citenamefont
  {Tsymbal}}]{PhysRevB.72.140404}%
  \BibitemOpen
  \bibfield  {author} {\bibinfo {author} {\bibfnamefont {K.~D.}\ \bibnamefont
  {Belashchenko}}, \bibinfo {author} {\bibfnamefont {J.}~\bibnamefont {Velev}},
  \ and\ \bibinfo {author} {\bibfnamefont {E.~Y.}\ \bibnamefont {Tsymbal}},\
  }\href {\doibase 10.1103/PhysRevB.72.140404} {\bibfield  {journal} {\bibinfo
  {journal} {Phys. Rev. B}\ }\textbf {\bibinfo {volume} {72}},\ \bibinfo
  {pages} {140404} (\bibinfo {year} {2005})}\BibitemShut {NoStop}%
\bibitem [{\citenamefont {Schubert}\ \emph {et~al.}(2012)\citenamefont
  {Schubert}, \citenamefont {Fehske}, \citenamefont {Fritz},\ and\
  \citenamefont {Vojta}}]{PhysRevB.85.201105}%
  \BibitemOpen
  \bibfield  {author} {\bibinfo {author} {\bibfnamefont {G.}~\bibnamefont
  {Schubert}}, \bibinfo {author} {\bibfnamefont {H.}~\bibnamefont {Fehske}},
  \bibinfo {author} {\bibfnamefont {L.}~\bibnamefont {Fritz}}, \ and\ \bibinfo
  {author} {\bibfnamefont {M.}~\bibnamefont {Vojta}},\ }\href {\doibase
  10.1103/PhysRevB.85.201105} {\bibfield  {journal} {\bibinfo  {journal} {Phys.
  Rev. B}\ }\textbf {\bibinfo {volume} {85}},\ \bibinfo {pages} {201105}
  (\bibinfo {year} {2012})}\BibitemShut {NoStop}%
\end{thebibliography}%

\end{document}